\documentclass[10pt]{amsart}
\usepackage{amssymb}
\usepackage{graphicx}
 \usepackage{lscape}
 \usepackage[all,cmtip]{xy}
 \usepackage{cancel}

\newtheorem{theorem}{Theorem}[section]
\newtheorem{lemma}[theorem]{Lemma}
\newtheorem{cor}[theorem]{Corollary}
\newtheorem{prop}{Proposition}
\theoremstyle{definition}
\newtheorem{definition}[theorem]{Definition}
\newtheorem{example}[theorem]{Example}

\theoremstyle{remark}
\newtheorem{remark}[theorem]{Remark}
\numberwithin{equation}{section}

\newcommand\gl{\mathfrak{gl}}
 \def\sl{\mathfrak{sl}}
 \newcommand\so{\mathfrak{so}}
 \newcommand\su{\mathfrak{su}}
 \newcommand\sa{\mathfrak{sa}}
 \newcommand\g{{\mathfrak g}}
 \newcommand\h{{\mathfrak h}}
 \def\k{{\mathfrak k}}
 \newcommand\s{{\mathfrak s}}
 
 \newcommand\n{{\mathfrak n}}
 
 \newcommand\solv{\mathfrak{solv}}

 \newcommand\R{{\mathbb R}}
 \newcommand\C{{\mathbb C}}
 \def\H{{\mathbb H}}

 \def\l{\left}
 \def\r{\right}

 \newcommand\ra{\rightarrow}
 \newcommand\lra{\longrightarrow}
 \newcommand\xynor{\ar@{->}}

 \newcommand\st{\,|\,}
 \newcommand\Lieder[1]{\mathcal{L}_{#1}}
 \newcommand\qbox[1]{\quad\mbox{#1}\quad}

 \newcommand\homo{\lambda}
 \newcommand\lagrangian{L}
 \newcommand\Ver{\mathcal{V}}
 \newcommand\Hor[1]{\mathcal{H}_{#1}}
 \newcommand\Wang{W}
 \newcommand\tWang{{\tilde{\Wang}}}
 \def\im{{\rm im}}

 \def\FR#1{v_{#1}}
 \def\parder#1{\frac{\partial}{\partial #1}}
 \newcommand\PFB[2]{[#1,#2]}

\begin{document}

% \title[short text for running head]{full title}
\title[Invariant Yang--Mills connections over non-reductive spaces]{Invariant Yang--Mills connections over Non-Reductive Pseudo-Riemannian Homogeneous Spaces}

% \author[short version for running head]{name for top of paper}
\author{Dennis The}
\address{Department of Mathematics \& Statistics.
McGill University, 805 Sherbrooke Street West, Montreal, QC, Canada, H3A 2K6}
%\curraddr{}
\email{dthe@math.mcgill.ca}
\thanks{The author was supported in part by an NSERC CGS-D and a Quebec FQRNT Fellowship.}

%    \subjclass is required.

\date{October 9, 2007}


 \begin{abstract}
We study invariant gauge fields over the 4-dimensional non-\linebreak{}reductive pseudo-Riemannian homogeneous spaces $G/K$ recently classified by Fels \& Renner (2006).  Given $H$ compact semi-simple, classification results are obtained for principal $H$-bundles over $G/K$ admitting: (1) a $G$-action (by bundle automorphisms) projecting to left multiplication on the base, and (2) at least one $G$-invariant connection.  There are two cases which admit nontrivial examples of such bundles and all $G$-invariant connections on these bundles are Yang--Mills.  The validity of the principle of symmetric criticality (PSC) is investigated in the context of the bundle of connections and is shown to fail for all but one of the Fels--Renner cases.  This failure arises from degeneracy of the scalar product on pseudo-tensorial forms restricted to the space of symmetric variations of an invariant connection.  In the exceptional case where PSC is valid, there is a unique $G$-invariant connection which is moreover universal, i.e. it is the solution of the Euler--Lagrange equations associated to any $G$-invariant Lagrangian on the bundle of connections.  This solution is a canonical connection associated with a weaker notion of reductivity which we introduce.
 \end{abstract}

 \maketitle

 \section{Introduction}
 \label{intro}

 Symmetry reduction methods have been extremely useful in the search for exact solutions of many of the partial differential equations (PDE) arising in mathematical physics.  By looking at the action of a particular group $G$ and restricting one's interest to the space of $G$-invariant fields, there is often a significant simplification in the field equations through a reduction in the number of independent or dependent variables (or both).  For example, in the case that the group acts transitively, all PDE reduce to algebraic equations; in the cohomogeneity one case, all PDE reduce to ordinary differential equations (ODE).  A famous instance of the use of symmetry reduction is the derivation of the Schwarzschild solution in General Relativity, which is static and spherically symmetric.  In this case, the Einstein field equations reduce to ODE that can be integrated exactly.  In a series of papers written by Coquereaux, Jadczyk \& Pilch \cite{CJ1983,JadEYM1984,JP1984,CJ1985,CJ1986,CJbook1988}, a symmetry reduction scheme for metrics and connections was established and used with much success.  However, they make the standing assumption that the symmetry group $G$ is compact, which in particular implies that the group orbits are {\em reductive}.  The resulting decompositions used in their reduction scheme depend on this reductivity.  Moreover, while the reduction of metrics is clear in Riemannian signature, problems with degeneracy occur in pseudo-Riemannian signature if the Killing vectors (i.e.\ infinitesimal symmetry generators) are null vectors.

  A homogeneous space (or orbit) $G/K$ is {\em reductive} if the Lie algebra $\k$ of $K$ admits an $Ad(K)$-invariant vector space complement $\s$ in the Lie algebra $\g$ of $G$, i.e.\ $\g = \k \oplus \s$.  For reductive homogeneous spaces admitting a $G$-invariant pseudo-Riemannian metric, the curvature tensor assumes a simple form \cite{KN-FDG1} and so the geometry of these spaces has been well-studied.  In the case of gauge theory over a reductive base manifold $G/K$, there is a canonical connection on the principal $K$-bundle $G \ra G/K$, which induces a corresponding canonical connection on any principal $H$-bundle $P \ra G/K$.  Important examples of reductive homogeneous spaces are the symmetric spaces.  For principal $H$-bundles over Riemannian symmetric spaces $G/K$, Harnad et al.\ \cite{HTS1980} established that the canonical connection satisfies the Yang--Mills equations.  Over general non-reductive homogeneous spaces, no notion of canonical connection is apparent.  However, we will introduce a weaker notion of reductivity (see Definition \ref{k0-reductive}) which leads to a notion of canonical connections on principal bundles over certain non-reductive homogeneous spaces (see Lemma \ref{lambda-reductive-lemma}).

 Essentially {\em all} applications of symmetry reduction to the study of PDE have involved groups acting with reductive orbits.  This is because the main symmetry groups of interest preserve some metric, and while it is easy to construct examples of non-reductive homogeneous spaces $G/K$, it is significantly more difficult to construct non-reductive examples which moreover admit a $G$-invariant metric.  The only systematic attempt so far to explicitly classify such spaces in low dimensions has been recent work in 2006 by Fels \& Renner \cite{FelsRenner2006}.  In this paper, all non-reductive homogeneous spaces $G/K$ up to dimension 4 admitting a $G$-invariant metric were classified.  They furthermore found that such spaces are necessarily: (1) non-Riemannian, and (2) essentially of dimension 4 or higher.  The first assertion is easy to establish: if $G$ is the isometry group of a $G$-invariant Riemannian metric on $G/K$, then $Ad(K)$ is compact and so the orthogonal complement of the subalgebra $\k$ (which is $Ad(K)$-invariant) with respect to the induced $Ad(K)$-invariant (positive-definite) inner product on $\g$ is an $Ad(K)$-invariant complementary subspace.  In non-Riemannian signature, the complication is that the ``perp'' of a given subspace $U$ does {\em not} in general produce a complementary subspace $V$, i.e.\ in general, $U \cap V \neq 0$.  The fact that there are no 2-dimensional examples and essentially (i.e.\ if $K$ is connected) no 3-dimensional examples is more difficult to establish and is one of the results of the Fels \& Renner analysis.  They discovered 8 classes of non-reductive pseudo-Riemannian homogeneous spaces in dimension 4 and furthermore classified all invariant Einstein metrics on these spaces.

 The purpose of this paper is to initiate a study of invariant gauge fields over non-reductive pseudo-Riemannian spaces using the Fels \& Renner classification as our starting point.  For any compact semi-simple structure group $H$, we classify all principal fibre bundles (PFB) over these spaces $G/K$ which admit: (1) a $G$-action (by bundle automorphisms) projecting to left multiplication on the base, and (2) at least one $G$-invariant connection.  We refer to such bundles as homogeneous PFB.  For such bundles admitting a $G$-invariant connection, a classification theorem is proved (Theorem \ref{classification-thm}).  Most of these bundles are necessarily trivial: of the 8 classes, only 2 classes admit nontrivial homogeneous PFB supporting $G$-invariant connections, namely:
 \begin{itemize}
 \item A5: $\g = A^1_{4,9} \rtimes \sl(2,\R)$ (7-dim.), and $\k \cong \mbox{Bianchi V}$ (3-dim.).  Here, $A^1_{4,9}$ is the 4-dim.\ solvable Lie algebra $\langle w_1,...,w_4 \rangle$ with
    \begin{align*}
      [w_1,w_4]=2w_1, \quad
      [w_2,w_3]=w_1, \quad
      [w_2,w_4]=w_2, \quad
      [w_3,w_4]=w_3.
    \end{align*}
 \item B3: $\g = \sa(2,\R) \times \R$ (6-dim.), and $\k \cong \R^2$.   (Here, $\sa(2,\R)$= special affine.)
 \end{itemize}
 We state our main result (see Theorems \ref{classification-thm} and \ref{YM-thm}) here:
 \begin{theorem}
 Given any homogeneous PFB with compact semi-simple structure group $H$ over a 4-dim.\ non-reductive pseudo-Riemannian space $G/K$ (with $K$ connected) of class A5 or B3, {\em all} $G$-inv.\ connections are Yang--Mills.  For A5, there is a {\em unique} $G$-inv.\ connection which is flat.  For B3, there exist non-flat connections.
 \end{theorem}

 These Yang--Mills connections are given explicitly in a local gauge on some global models $G/K$ in Sections \ref{B3-section} and \ref{A5-section}.
 
 Given a $G$-invariant metric and volume form, the Yang--Mills equations are $G$-invariant equations which are the Euler--Lagrange equations of a $G$-invariant Lagrangian.  The existence of $G$-invariant Yang--Mills connections leads us naturally to investigate the validity of the principle of symmetric criticality (PSC) for the bundle of connections.  In its original formulation due to Palais \cite{Palais1979}, PSC states that given an invariant functional, critical points of the symmetry-reduced functional (i.e.\ critical points along symmetric variations) correspond to (symmetric) critical points (i.e.\ with respects to arbitrary variations) of the original invariant functional.  This is sometimes dubbed: ``critical symmetric points'' are ``symmetric critical points''.  In the context of the calculus of variations, it becomes restrictive to consider the Lagrangian functional since this involves an integration over the base manifold which may not be compact and hence certain decay conditions at infinity may need to be introduced.  (In fact, in all of our non-reductive examples, the base manifold is non-compact.)  Consequently, we instead work with a local formulation of PSC due to Anderson, Fels \& Torre \cite{AndersonFels1997,AFT2001} stated in terms of Lagrangian forms (instead of functionals) defined in the context of the variational bicomplex \cite{Anderson-VB,Anderson1992}.  In this context, given a $G$-action on a bundle, PSC states that for {\em any} $G$-invariant Lagrangian form defined on (jets of) the given bundle, local solutions of the Euler--Lagrange equations of the symmetry-reduced Lagrangian correspond to local invariant solutions of the Euler--Lagrange equations of the original Lagrangian.  The bundle of connections over these non-reductive spaces provide interesting examples where this natural principle generally {\em fails}.  We show that the failure is a consequence of the degeneracy of the scalar product on pseudo-tensorial forms restricted to the space of symmetric variations of an invariant connection.  The one instance among the examples where PSC does hold is in the A5 case.  The unique $G$-invariant connection in the A5 case is (by PSC) an example of a {\em universal} connection - i.e.\ it is the solution of the Euler--Lagrange equations associated to {\em any} $G$-invariant Lagrangian defined on (jets of) the bundle of connections.

 \section{Gauge theory on homogeneous principal fibre bundles}
 \label{gauge-theory}

 In this section, as preparation for our investigation of the Yang--Mills equations on non-reductive spaces, we revisit the reduction of the Yang--Mills equations over homogeneous spaces.

A {\em homogeneous} principal fibre bundle (PFB) is a PFB $P=P(G/K,H)$ over a homogeneous space together with a $G$-action by PFB automorphisms which projects to left multiplication of $G$ on the base manifold $G/K$ and is compatible with the right action of the structure group $H$.  A homogeneous PFB is itself a homogeneous space. Consequently, by using some familiar results due to Chevalley and Eilenberg \cite{ChevEilen1948}, the $G$-invariant objects relevant in gauge theory (i.e.\ connections, curvature, and pseudo-tensorial forms in general) on these homogeneous PFB have corresponding analogues in a purely Lie algebraic setting.  We describe in detail this mapping between $G$-invariant pseudo-tensorial forms and the corresponding Lie algebraic data.  In particular, we recover Wang's theorem which parametrizes the spaces of $G$-invariant connections by certain $K$-invariant linear maps between $\g$ and $\h$.
The mapping moreover allows us to naturally define an exterior covariant calculus and write the corresponding reduced Yang--Mills equations algebraically.  The solution of these algebraic equations are in 1-1 correspondence with the $G$-invariant solutions of the original Yang--Mills equations.

While the Yang--Mills equations have of course been investigated over homogeneous spaces (see for example \cite{HTS1980,Laquer1984,Gotay1989,Koiso1990,Darian1997,DK1997}), our derivation will make no mention of the compactness of the isotropy group $K$, nor depend on the reductivity of $(\g,\k)$, which is a prevalent assumption in the literature.  Moreover, we believe that the point of view taken in our approach to be new, i.e.\ (1) appealing to results of Chevalley \& Eilenberg to transfer geometric objects and operators into an algebraic setting, and (2) defining an exterior covariant calculus on $\Lambda^*(\g,K;\h)$, the space of $\h$-valued chains on $\g$ which are $K$-basic.  (Note the $K$-invariance here is defined with respect to the $K$-action given in \eqref{K-action} which twists an action on $\g$ and $\h$.)

 For any given Lie group $G$, we will always use the convention that the corresponding Lie algebra $\g$ is identified with the tangent space $T_e G$ at the identity as well as the {\em left}-invariant vector fields on $G$.

 \subsection{Homogeneous principal bundles}

 We recall the following explicit classification of homogeneous PFB due to Harnad, Shnider and Vinet \cite{HSV1980}.

 \begin{theorem}
   There exists a 1-1 correspondence between:
   \begin{enumerate}
   \item homogeneous PFB $P(G/K,H)$ (modulo bundle equivalence), and
   \item group homomorphisms $\homo : K \ra H$ (modulo conjugation in $H$).
   \end{enumerate}
 \end{theorem}

 Given a homomorphism $\homo : K \ra H$, let $\PFB{g}{h}$ denote the equivalence class of $G\times H$ defined by $(g,h) \sim (gk,\homo(k)^{-1}h)$, $\forall k \in K$.  The corresponding bundle is
 \begin{align*}
 P_\homo = G \times_\homo H = (G\times H)/\sim
 \end{align*}
 with projection $\pi: P_\homo \ra G/K$ given by $\PFB{g}{h} \mapsto \overline{g}$.
 We have a transitive {\em left} action of $G \times H$ on $P_\homo$ via
 \begin{align}
    \psi_{(g,h)}\l(\PFB{g'}{h'}\r) = \PFB{gg'}{h'h^{-1}},
    \label{GH-Plambda-action}
 \end{align}
 i.e.\ $\psi_{(g,h)} = L_g R_{h^{-1}}$, where $L_g$ and $R_h$ denote the left $G$-action and right $H$-action on $P_\homo$ respectively.  Thus, $P_\homo$ is itself a homogeneous space.  Noting that the stabilizer for this action at the point $\PFB{e}{e}$ is
 \begin{align*}
   \hat{K} = Stab_{G\times H}(\PFB{e}{e}) = \{ (g,h) \st \PFB{g}{h^{-1}} = \PFB{e}{e} \}
   = \{ (k,\homo(k)) \st k \in K \} \cong K,
 \end{align*}
 we have a bijection (in fact, diffeomorphism)
 \begin{align}
   \Psi : (G \times H) / \hat{K} \stackrel{\cong}{\lra} P_\homo,
   \quad\quad \overline{(g,h)} \mapsto \PFB{g}{h^{-1}}
   \label{Psi-bijection}
 \end{align}
 which respects the left $(G\times H)$-action on either space:
 \begin{align*}
   \xymatrix{
   (G \times H) / \hat{K}
   \xynor[rr]^{\Psi} \xynor[d]_{L_{(g,h)} = L_g L_h} && P_\homo \xynor[d]^{\psi_{(g,h)} = L_g R_{h^{-1}}}\\
   (G \times H) / \hat{K} \xynor[rr]^{\Psi} && P_\homo
   }
 \end{align*}
 where $L_{(g,h)} = L_g L_h$ denotes the usual left action of $G\times H$ on itself and also on the quotient $(G \times H)/\hat{K}$.  Consequently, we can identify
 \begin{align}
   \Psi_* : (\g\times\h)/\hat{\k} &\stackrel{\cong}{\lra} T_p P_\homo, \quad
   \overline{(x,y)} \mapsto (x,-y)^*_p := \l.\frac{d}{dt}\r|_{t=0} L_{exp(tx)} R_{exp(-ty)}(p),
   \label{TPlambda}
 \end{align}
 where $p=\PFB{e}{e} \in P_\homo$ and
 \begin{align*}
 \hat{\k} = \{ (x,\homo_*(x) \st x \in \k \}.
 \end{align*}
 We will abbreviate the vertical vector $(0,y)^*_p \in T_p P_\homo$ by $y^*_p$.
 Note $K$ acts on $T_p P_\homo$ via $L_{k*} R_{\homo(k)^{-1}*}$, $K$ acts on $(\g \times \h) / \hat{\k}$ via
 \begin{align*}
   k \cdot \overline{(x,y)} = \overline{(Ad_k(x), Ad_{\homo(k)}(y))},
 \end{align*}
 and the isomorphism \eqref{TPlambda} respects the $K$-action on both spaces.

 \subsection{Invariant pseudo-tensorial forms on $P_\homo$}

 By definition, an $\h$-valued form on $P_\homo$ is {\em equivariant} with respect to the left action $\psi$ of $G\times H$ on $P_\homo$ given in \eqref{GH-Plambda-action} and the representation $\rho : G \times H \ra Aut(\h)$, $(g,h) \mapsto Ad_h$ iff for all $(g,h) \in G \times H$,
 \begin{align*}
   \psi_{(g,h)}^* \omega = \rho_{(g,h)}  \omega, \qbox{i.e.} L_g^* \omega = \omega, \quad R_h^*\omega = Ad_{h^{-1}} \omega.
 \end{align*}
 That is, the $G$-invariant pseudo-tensorial forms on $P_\homo$, denoted $\Omega^*_{pseudo.}(P_\homo;\h)^G$, are precisely the equivariant forms on $P_\homo$, denoted $\Omega^*_{equiv.}(P_\homo;\h)$.  Using $\Psi$ as in equation \eqref{Psi-bijection}, we have an isomorphism $\Omega^*_{equiv.}(P_\homo;\h) \cong \Omega^*_{equiv}((G\times H)/\hat{K};\h)$ given by $\Psi^*$.  Chevalley \& Eilenberg investigated equivariant forms on homogeneous spaces, and a particular case of their Thm.13.1 in \cite{ChevEilen1948} yields the isomorphism
 \begin{align*}
 \Omega^*_{equiv}((G\times H)/\hat{K};\h) \cong \Lambda^*(\g \times \h, \hat{K};\h),
 \end{align*}
 where $\Lambda^*(\g \times \h, \hat{K};\h) \subset \Lambda^*(\g \times \h;\h)$ denotes the $\h$-valued chains on $\g\times\h$ which are $\hat{K}$-basic, i.e.\ they vanish on $\hat{\k}$ and are $\hat{K}$-invariant.  We will define the $\hat{K}$-invariance condition more precisely below.  Let us denote by $\tilde\Phi$ the linear isomorphism given by the composition of the above isomorphisms:
 \begin{align}
   \tilde\Phi : \Omega^*_{pseudo.}(P_\homo;\h)^G \lra \Lambda^*(\g \times \h, \hat{K};\h), \quad\quad
    \omega \mapsto (\pi^*\Psi^*\omega)_{(e,e)},
  \label{localization}
 \end{align}
 where by a small abuse of notation we use $\pi$ to also denote $\pi : G\times H \ra (G\times H)/\hat{K}$.

 Let $\varphi^i \in \Lambda^i(\g\times\h;\h)$ and let $z,z_i \in \g\times \h$.  We have the graded commutator and interior product
 \begin{align*}
   &[\varphi^p,\varphi^q](z_1, ..., z_{p+q})\\
   & \qquad = \frac{1}{p!q!} \sum_{\sigma \in S_{p+q}} (-1)^{sgn\,\sigma} [\varphi^p(z_{\sigma(1)},...,z_{\sigma(p)}),\varphi^q(z_{\sigma(p+1)},...,z_{\sigma(p+q)})], \nonumber\\
&   (i_z\varphi^p)(z_1,...,z_{p-1}) = \varphi^p(z,z_1,...,z_{p-1}),
 \end{align*}
 and with respect to representation $\rho_*(z) = ad_{proj_\h(z)}$, we define the exterior derivative and Lie derivative
 \begin{align}
   (d\varphi^p)(z_1,...,z_{p+1}) &=
    \sum_{i=1}^{p+1} (-1)^{i-1} [proj_\h(z_i),\varphi^p(z_1,...,\hat{z}_i,...,z_{p+1})] \label{extd}\\
    &\quad\quad + \sum_{i < j} (-1)^{i+j} \varphi^p([z_i,z_j],z_1,...,\hat{z}_i,...,\hat{z}_j,...,z_{p+1}), \nonumber\\
    (\Lieder{z}\varphi^p)(z_1,...,z_p) &= [proj_\h(z),\varphi^p(z_1,...,z_p)]
    - \sum_{i=1}^p \varphi^p(z_1,...,[z,z_i],...,z_p). \label{Lie-der}
 \end{align}
We have the usual properties \cite{ChevEilen1948}, \cite{HochSerre1953}:
 \begin{itemize}
 \item  $[\varphi^p,\varphi^q]=-(-1)^{pq}[\varphi^q,\varphi^p]$
 \item  $(-1)^{pr}[[\varphi^p,\varphi^q],\varphi^r] + (-1)^{qp}[[\varphi^q,\varphi^r],\varphi^p]
            + (-1)^{rq}[[\varphi^r,\varphi^p],\varphi^q] = 0$
 \item $d^2=0$,  \quad\quad $i_z^2=0$, \quad\quad $i_{z_1} i_{z_2} = - i_{z_2} i_{z_1}$,
        \quad\quad $\Lieder{[z_1,z_2]} = [\Lieder{z_1}, \Lieder{z_2}]$
 \item $[\Lieder{z_1},i_{z_2}]=[i_{z_1},\Lieder{z_2}]=i_{[z_1,z_2]}$, \quad\quad $d \circ \Lieder{z} = \Lieder{z} \circ d$
 \item Cartan's identity: $\Lieder{z} = i_z \circ d + d\circ i_z$
 \item $d, i_z, \Lieder{z}$ are derivations of $\Lambda^*(\g\times\h;\h)$ of degree $+1,-1,0$ respectively.
 \end{itemize}
 In particular, the space $\Lambda^*(\g\times\h;\h)$ is a differential graded algebra (DGA) with respect to the exterior derivative and the graded commutator.

 The subgroup $\hat{K}$ acts on $\h$ via the representation $\rho$, and $\hat{K}$ acts on $\g\times \h$ via the adjoint representation.  This induces a corresponding action on $\Lambda^*(\g\times\h;\h)$.  If $\hat{k} = (k,\homo(k))$ and $z=(x,y)$, then
 \begin{align*}
 (\hat{k} \cdot \varphi^p)(z_1,...,z_p)
    = Ad_{\homo(k)}(\varphi^p(\hat{k}^{-1} \cdot z_1, ..., \hat{k}^{-1} \cdot z_p )),
 \end{align*}
 and
 \begin{align*}
   \hat{k} \cdot z = \hat{k} \cdot (x,y) = (Ad_k(x), Ad_{\homo(k)}(y)).
 \end{align*}
 The corresponding infinitesimal action of $\hat{K}$ on $\Lambda^*(\g\times\h;\h)$ is given naturally through the Lie derivative, and as usual if $\hat{K}$ is connected, then $\hat{K}$-invariance of forms (i.e.\ $\hat{k} \cdot \varphi = \varphi$ for all $\hat{k} \in \hat{K}$) is equivalent to $\hat{\k}$-invariance of forms (i.e.\ $\Lieder{z} \varphi = 0$ for all $z \in \hat{\k}$).  A form is $\hat{K}$-semibasic if it vanishes on $\hat{\k}$, i.e.\ $i_z \varphi = 0$ for all $z \in \hat{\k}$.  The subspace $\Lambda^*(\g\times\h,\hat{K};\h)$ is the set of $\hat{K}$-basic (i.e.\ $\hat{K}$-semibasic and $\hat{K}$-invariant) chains in $\Lambda^*(\g\times\h;\h)$.  Using the properties above, the exterior derivative and graded commutator on $\Lambda^*(\g\times\h;\h)$ naturally restrict to $\Lambda^*(\g\times\h,\hat{K};\h)$ making the latter a DGA, and the map $\tilde\Phi$ is a DGA isomorphism.

 \subsection{Wang's theorem and canonical connections}

 Recall that a connection (or gauge field) on a PFB $P=P(M,H)$ is by definition a pseudo-tensorial $\h$-valued 1-form $\omega \in \Omega^1_{pseudo.}(P;\h)$ which satisfies $\omega( y^* ) = y$, where $y^*$ is the fundamental vertical vector field induced by $y \in \h$.  Using the correspondence $\tilde\Phi$ given in \eqref{localization}, we immediately recover Wang's theorem \cite{Wang1958}, which parametrizes the space of $G$-invariant connections on a homogeneous PFB $P_\homo(G/K,H)$.
  \begin{theorem}[Wang]  \label{Wang-thm}
 The $G$-invariant connections on $P_\homo(G/K,H)$ are in 1-1 correspondence with linear maps $\Wang : \g \ra \h$ such that:
 \begin{enumerate}
   \item $\Wang(x) = \homo_*(x), \quad \forall x \in \k$,
   \item $\Wang(Ad_k v) = Ad_{\homo(k)}(\Wang(v)), \quad \forall v \in \g,\ \forall k \in K$.
 \end{enumerate}
 Such maps will be referred to as {\em Wang maps}.
 \end{theorem}

 \begin{proof} We describe the image of the set of $G$-invariant connections under the correspondence $\tilde\Phi$.  Let $\omega$ be a $G$-invariant connection and let $\tWang = \tilde\Phi(\omega) \in \Lambda^1(\g\times\h,\hat{K};\h)$.  We have for any $y \in \h$, $\tWang(0,y) = (\Psi^*\omega)(\overline{(0,y)}) = \omega(-y^*_p) = -y$, where $p=\PFB{e}{e}$.
 Define $\Wang : \g \ra \h$ by $\Wang(x)=\tWang(x,0)$.  Then $\tWang(x,y) = \Wang(x) - y$ and so $\tWang$ is completely determined by $\Wang$.  Since $\tWang$ vanishes on $\hat{\k}$, then $0 = \tWang(x,\homo_*(x)) = \Wang(x)-\homo_*(x)$ for all $x \in \k$, i.e.\ $\Wang|_\k = \homo_*$.  Since $\tWang$ is $\hat{K}$-invariant, then $\Wang$ is $K$-invariant (with respect to $k\cdot \Wang = Ad_{\homo(k)} \circ \Wang \circ Ad_{k^{-1}}$).  Conversely, any such $\Wang$ defines $\tWang(x,y) = \Wang(x) - y$ and a $G$-invariant connection $\tilde\Phi^{-1}(\tWang)$.
 \end{proof}

 We note that it suffices to verify the second condition for Wang maps on a complement to $\k$ in $\g$, since the verification of the condition on $\k$ follows immediately from the identity $\homo \circ Ad_k = Ad_{\homo(k)} \circ \homo$ (which comes from the fact that $\homo$ is a homomorphism).

 The $G$-invariant connection on $P_\homo$ corresponding to a Wang map $\Wang$ is
 \begin{align*}
   \omega_{\PFB{g}{h}} = Ad_{h^{-1}} \circ \Wang \circ (\Theta_G)_g + (\Theta'_H)_h,
 \end{align*}
 where $\Theta_G$ is the left-invariant Maurer--Cartan form on $G$ and $\Theta'_H$ is the right-invariant  Maurer--Cartan form on $H$.  In particular, at $p = \PFB{e}{e}$,
 \begin{align}
   \omega_p((x,y)^*_p) = W((\Theta_G)_e(x)) + (\Theta'_H)_e(y) = W(x) + y.
 \label{omega-W}
 \end{align}

 \begin{cor} The principal $K$-bundle $G \ra G/K$ admits a $G$-invariant connection iff $(\g,\k)$ is reductive \cite{KN-FDG1}.
 \end{cor}
 \begin{proof} 
Here, $\homo = id_K$.  Given a Wang map $\Wang : \g \ra \k$, let $\s = \ker \Wang$.  By the first Wang condition, we have $\Wang|_\k = id_\k$ so $\Wang$ is onto.  Also,  $dim(\g) = dim(\ker \Wang) + dim(\im \Wang) = dim(\s) + dim(\k)$ and so $\g=\k\oplus\s$.  Since $\Wang$ is $K$-invariant, then $\s$ is $Ad(K)$-invariant.  Conversely, if $(\g,\k)$ is reductive with reductive decomposition $\g = \k \oplus \s$, then $\omega = proj_\k(\Theta_\g)$ is the {\em canonical} $G$-invariant connection on $G \ra G/K$.  It has corresponding Wang map
 \begin{align*}
 \Wang(x) = \l\{ \begin{array}{cc} x, & x \in \k\\
                        0, & x \in \s \end{array} \r. .
 \end{align*}
 \end{proof}

 For $G/K$ reductive, the canonical connection on $G\ra G/K$ induces via the map $f(g) = \PFB{g}{e}$ a corresponding canonical connection on any $P_\homo(G/K,H)$ \cite{HTS1980}.  We now define a weaker notion of reductivity.

 \begin{definition} \label{k0-reductive} Let $\k_0 \subset \k$ be an ideal.  We refer to $G/K$ (or $(\g,\k)$) as {\em $\k_0$-reductive} if there exists a subspace $\s \subset \g$ with $\g = \k \oplus \s$ and for any $k \in K$,
 \begin{align*}
    Ad_k(\s) \subset \k_0 \oplus \s.
 \end{align*}
 Given a homomorphism $\homo : K \ra H$, let us also say that $G/K$ is {\em $\homo$-reductive} if it is $\ker(\homo_*)$-reductive.
 \end{definition}

 Note that the usual definition of reductivity is the same as $0$-reductivity and that any $(\g,\k)$ is clearly $\k$-reductive.  The $\homo$-reductivity property leads to the existence of canonical connections on $P_\homo(G/K,H)$.

 \begin{lemma} \label{lambda-reductive-lemma} Let $\homo : K \ra H$ be a homomorphism and suppose that $G/K$ is $\homo$-reductive with decomposition $\g = \k \oplus \s$.
 Then there exists a canonical $G$-invariant connection on the homogeneous PFB $P_\homo = P_\homo(G/K,H)$ associated to the Wang map
 \begin{align*}
   \Wang_\homo(x) = \l\{ \begin{array}{cl} \homo_*(x), & x \in \k\\
                        0, & x \in \s
                        \end{array} \r. .
 \end{align*}
 In particular, any reductive $G/K$ is $\homo$-reductive, and we recover the canonical connection mentioned prior to Definition \ref{k0-reductive}.
 \end{lemma}

 \begin{proof}
 It suffices to verify the second condition for Wang maps (as in Theorem \ref{Wang-thm}) on $\s$.  If $v \in \s$, then by definition $W_\homo(v)=0$, and since $Ad_k(v) \in \ker(\homo_*) \oplus \s$ for any $k\in K$, then $W(Ad_k(v))=0$.
 \end{proof}

 \begin{example} If $\homo_0 : K \ra H$ is the trivial homomorphism, then $G/K$ is obviously $\homo_0$-reductive and 
 $\Wang_{\homo_0}$ is the canonical flat connection on $P_{\homo_0}(G/K,H) \cong G/K \times H$.
 \end{example}

 \subsection{Covariant calculus on $\Lambda^*(\g\times\h;\h)$}

 A connection on $P \ra M$ is equivalently defined as a smooth (right) $H$-invariant horizontal distribution $\{H_p \}_{p \in P}$ on $P$ which is pointwise complementary to the vertical distribution $\{V_p \}_{p \in P}$.  This defines horizontal and vertical projection operators, leading to the notion of horizontal lift of a vector field, an exterior covariant calculus, etc.  We discuss the corresponding algebraic analogues in this section.

 Let us fix {\em any} linear map $\Wang : \g\ra\h$, and define $\tWang : \g\times\h \ra \h$ by $\tWang(x,y) = \Wang(x)-y$.  Define the vertical and horizontal projection operators on $\g\times\h$,
 \begin{align*}
    \Ver(x,y) \mapsto (0,y-\Wang(x)), \quad\quad  \Hor{\tWang}(x,y) \mapsto (x,\Wang(x)).
 \end{align*}
 A vector $z \in \g\times\h$ is {\em horizontal} if $\Hor{\tWang}(z)=z$ (iff $\Ver(z)=0$ iff $\tWang(z)=0$) and it is {\em vertical} if $\Ver(z)=z$ (iff $\Hor{\tWang}(z)=0$). Any vector in $\g\times\h$ decomposes uniquely into horizontal and vertical parts, and the commutator of horizontal and vertical vectors is vertical:
 \begin{align*}
   [(x,\Wang(x)),(0,y)] = (0,[\Wang(x),y]).
 \end{align*}

 \begin{remark}
 This differs from the corresponding notion on a PFB.  There, given a connection, the horizontal lift of any vector field on the base manifold commutes with the fundamental vertical vector fields.
 \end{remark}

 Let $\Lambda^*_\h(\g\times\h;\h)$ be the subspace of $\h$-semibasic forms, i.e.\ $i_z\varphi=0$, $\forall z\in\h$.  In general, $d$ does not preserve $\Lambda^*_\h(\g\times\h;\h)$.  The {\em exterior covariant derivative} with respect to $\tWang$ is
  \begin{align*}
  d_\tWang : \Lambda^*(\g\times\h;\h) \ra \Lambda^*_\h(\g\times\h;\h), \quad\quad \varphi \mapsto  d_\tWang \varphi := d\varphi \circ \Hor{\tWang}
 \end{align*}
 Define the {\em curvature} of $\tWang$ to be
 \begin{align*}
   F_\tWang = d_\tWang \tWang \in \Lambda^2_\h(\g\times\h;\h).
 \end{align*}

 \begin{prop} We have the following identities:
 \begin{enumerate}
 \item $F_\tWang(z_1,z_2) = [\Wang(x_1),\Wang(x_2)] - \Wang([x_1,x_2])$, where $z_i = (x_i,y_i)$.  Thus, $F_\tWang = 0$ iff $\Wang$ is a Lie algebra homomorphism.
 \item $F_{\tWang} = d\tWang + \frac{1}{2} [\tWang,\tWang]$ and $d_\tWang F_\tWang=0$
 \item For $\varphi \in \Lambda^*_\h(\g\times\h;\h)$, $d_\tWang \varphi = d\varphi + [\tWang,\varphi]$ and $d_\tWang^2\varphi = [F_\tWang,\varphi]$.
 \item If $\Wang : \g \ra \h$ is a Wang map, then
 \begin{itemize}
 \item $\Hor{\tWang}$ commutes with the $\hat{K}$-action on $\g\times\h$.
 \item $d_\tWang$ preserves $\Lambda^*(\g\times\h,\hat{K};\h)$ and $\Lambda^*_\h(\g\times\h,\hat{K};\h)$.  In particular, $F_\tWang \in \Lambda^2_\h(\g\times\h,\hat{K};\h)$.
 \end{itemize}
 \end{enumerate}
 \end{prop}

 The proof is straightforward.  These identities can be established immediately on $\Lambda^*(\g\times\h,\hat{K};\h)$ using the following result and corresponding identities on PFB.

 \begin{prop} The map $\tilde\Phi: \Omega^*_{pseudo.}(P_\homo;\h)^G \ra \Lambda^*(\g\times\h,\hat{K};\h)$ is a DGA isomorphism which commutes with the exterior covariant derivative, i.e.\ if $\omega$ is a $G$-invariant connection on $P_\homo$, $\tWang = \tilde\Phi(\omega)$, and $d_\omega$ and $d_{\tWang}$ are the exterior covariant derivatives on $\Omega^*(P_\homo;\h)$ and $\Lambda^*(\g\times\h,\hat{K};\h)$ respectively, then
 \begin{align*}
   d_{\tWang} \circ \tilde\Phi = \tilde\Phi \circ d_\omega.
 \end{align*}
 \end{prop}

 \begin{proof} The linear isomorphism $\tilde\Phi$ commutes with $d$ and the graded commutator, hence is a DGA isomorphism.  Given a connection $\omega$, let $\Hor{\omega}$ denote the projection onto the horizontal distribution determined by $\omega$. Identifying $z =(x,y) \in \g\times\h = T_{(e,e)}(G\times H)$ with its corresponding left-invariant vector field on $G\times H$, we have for $\varphi \in \Omega^r_{pseudo.}(P_\homo;\h)^G$,
 \begin{align*}
   (\tilde\Phi \, d_\omega \varphi)(z_1,...,z_{r+1})
   &= (d_\omega \varphi)(\Psi_*\pi_*(z_1),...,\Psi_*\pi_*(z_{r+1})) \\
   &= (d\varphi)(\Hor{\omega}(\Psi_*\pi_*(z_1)),...,\Hor{\omega}(\Psi_*\pi_*(z_{r+1}))) \\
   (d_{\tWang} \tilde\Phi \,\varphi)(z_1,...,z_{r+1}) &=
    d(\pi^*\Psi^*\varphi)(\Hor{\tWang}(z_1),...,\Hor{\tWang}(z_{r+1}))\\
    &= (\pi^*\Psi^*d\varphi)(\Hor{\tWang}(z_1),...,\Hor{\tWang}(z_{r+1}))\\
    &= (d\varphi)(\Psi_*\pi_*\Hor{\tWang}(z_1),...,\Psi_*\pi_*\Hor{\tWang}(z_{r+1})).
 \end{align*}
 It suffices to show that $\Psi_*\pi_* \circ \Hor{\tWang}(z) = \Hor{\omega} \circ \Psi_*\pi_*(z)$.
 We have
 \begin{align*}
   \Psi_*\pi_*(\Hor{\tWang}(z)) = \Psi_*\pi_*(x,\Wang(x)) = (x,-\Wang(x))^*_p,
 \end{align*}
 where $p = \PFB{e}{e}$. Since $\omega$ is $G$-invariant, then so is the horizontal distribution corresponding to $\omega$. Thus, it suffices to evaluate $\Hor{\omega} \circ \Psi_*\pi_*(z)$ at $p$.
We have that $\omega_p((x,-y)^*_p) = \Wang(x) - y$ identifies the generator for the vertical part of $(x,-y)^*_p$.
 \begin{align*}
   \Hor{\omega}(\Psi_*\pi_*(z)) &= \Hor{\omega}((x,-y)^*_p) = (x,-y)^*_p - (0,\omega_p((x,-y)^*_p))^*_p\\
   &= (x,-y)^*_p - (0,\Wang(x)-y)^*_p = (x,-W(x))^*_p
 \end{align*}
 Thus, the exterior covariant derivative commutes with $\tilde\Phi$.
 \end{proof}

 \begin{remark}
 A simple calculation shows that
 \begin{align*}
   \hat{k} \cdot (i_z \varphi) = i_{\hat{k} \cdot z} \varphi, \qbox{and}
   \hat{k} \cdot (\Lieder{z} \varphi) = \Lieder{\hat{k} \cdot z} \varphi
 \end{align*}
 so in general the interior product and Lie derivative do not preserve $\Lambda^*(\g\times\h,\hat{K};\h)$.
 \end{remark}

 Tensorial forms are those pseudo-tensorial forms which vanish on vertical vector fields.
 In general, $d$ does not restrict to tensorial forms, but since the graded commutator does and since $(0\times \h)/ \hat{\k} \cong V_p \subset T_p P_\homo$, then we have that:

 \begin{prop} The map $\tilde\Phi : \Omega^*_{tens.}(P_\homo;\h)^G \ra \Lambda^*_\h(\g\times\h,\hat{K};\h)$ is a graded algebra isomorphism which commutes with the exterior covariant derivative.
 \end{prop}

 We can identify $\Lambda^*_\h(\g\times\h,\hat{K};\h)$ with $\Lambda^*(\g,K;\h)$, the subspace of $K$-basic $\h$-valued chains on $\g$, and consider a notion of exterior covariant derivative $d_\Wang$ (associated to a Wang map $\Wang : \g \ra \h$).  However, there is no natural notion of ``horizontal'' subspace of $\g$  which can be used to define $d_\Wang$  in this new setting.

 \subsection{Covariant calculus on $\Lambda^*(\g;\h)$}

 In addition to the absence of the notion of a ``horizontal'' subspace, one also lacks a natural notion of exterior and Lie derivative on $\Lambda^*(\g;\h)$ which takes into account the given homomorphism $\homo : K \ra H$.  We can define the graded commutator and interior product as usual, but the definition of the exterior and Lie derivatives on $\Lambda^*(\g;\h)$ requires a representation of $\g$ on $\h$.  A map $\homo_* : \k \ra \h$ defines a representation of $\k \ra End(\h)$ via $x \mapsto ad_{\homo_*(x)}$, but for Wang maps $\Wang : \g \ra \h$ it is easy to show that:

 \begin{lemma} \label{L-rep} Let $\Wang : \g \ra \h$ be a Wang map and suppose that $center(\h) = 0$.  Then the map $\rho_\Wang : \g \ra End(\h)$, $x \mapsto ad_{\Wang(x)}$ is a Lie algebra representation iff $\Wang$ is a Lie algebra homomorphism.
 \end{lemma}

% \begin{proof} Let $F_\Wang(x_1,x_2) := \Wang([x_1,x_2]) - [\Wang(x_1),\Wang(x_2)]$.  Then $\rho_\Wang$ is a representation iff $ad_{\Wang([x_1,x_2])} = [ad_{\Wang(x_1)},ad_{\Wang(x_2)}] = ad_{[\Wang(x_1),\Wang(x_2)]}$ iff $ad_{F_\Wang(x_1,x_2)} = 0$ iff $F_\Wang(x_1,x_2) \in \ker(ad) = center(\h) = 0$.  Thus $F_\Wang=0$, i.e.\ $\Wang$ is a Lie algebra homomorphism.
% \end{proof}

We do not in general have a {\em canonical} nontrivial representation of $\g$ on $\h$ without specifying any additional structure.  (For example, specifying a complementary subspace $\s$ so that $\g = \k \oplus \s$ would yield a projection map onto $\k$ and hence a representation $x \mapsto ad_{\homo_*(proj_\k(x))}$.)  We define the exterior and Lie derivatives on $\Lambda^*(\g;\h)$ with respect to the {\em trivial} representation of $\g$ on $\h$.  For $\varphi \in \Lambda^p(\g;\h)$,
 \begin{align}
   (d\varphi)(x_1,...,x_{p+1}) &= \sum_{i < j} (-1)^{i+j} \varphi([x_i,x_j],x_1,...,\hat{x}_i,...,\hat{x}_j,...,x_{p+1}), \label{triv-d}\\
    (\Lieder{x}\varphi)(x_1,...,x_p) &= - \sum_{i=1}^p \varphi(x_1,...,[x,x_i],...,x_p). \label{triv-Lie}
 \end{align}
 These make $\Lambda^*(\g;\h)$ into a DGA with the usual identities holding (see identities following \ref{Lie-der}).

 We have a $K$-action on $\h$ via $Ad_{\homo(k)}$ and a $K$-action on $\g$ via the adjoint representation.
 This induces a representation on $\Lambda^*(\g;\h)$, namely for $x_i \in \g$,
 \begin{align}
   (k\cdot\varphi)(x_1,...,x_p) = Ad_{\homo(k)}(\varphi(Ad_{k^{-1}}(x_1),...,Ad_{k^{-1}}(x_p))).
 \label{K-action}
 \end{align}
 If $K$ is connected, then $K$-invariance (i.e.\ $k\cdot \varphi = \varphi$) is equivalent to $\k$-invariance.   Note $\k$-invariance is {\em not} simply $\Lieder{x}\varphi=0$ for all $x \in \k$.
 Instead it is the condition
 \begin{align*}
   \Lieder{x}^{(\homo)} \varphi =0, \quad \forall x \in \k,
 \end{align*}
 where
 \begin{align*}
   \Lieder{x}^{(\homo)}\varphi := \Lieder{x}\varphi + [\homo_*(x),\varphi].
 \end{align*}
 Now, $\Lambda^*(\g,K;\h)$ is the space of $\h$-valued $K$-basic (i.e.\ $K$-invariant, $K$-semibasic) chains on $\g$.  Note that in general Wang maps $\Wang$ are not elements of $\Lambda^1(\g,K;\h)$ since they are not $K$-semibasic.

 The projection $\pi_\g : \g\times\h \twoheadrightarrow \g$ induces an injection $\pi_\g^* : \Lambda^*(\g;\h) \hookrightarrow\Lambda^*(\g\times\h;\h)$ which maps onto $\Lambda^*_\h(\g\times\h;\h)$.
 Since $\pi_\g$ commutes with the $\hat{K}$-action (or $K$-action) on $\g\times\h$ and the $K$-action on $\g$, then $\pi_\g^*$ restricts to an isomorphism $\pi_\g^* : \Lambda^*(\g,K;\h) \hookrightarrow \Lambda^*_\h(\g\times\h,\hat{K};\h)$.
 We can circumvent the lack of a notion of horizontality by defining the exterior covariant derivative on $\Lambda^*(\g,K;\h)$ through
 \begin{align*}
   d_\Wang\varphi := (\pi_\g^*)^{-1}(d_\tWang(\pi_\g^*\varphi)).
 \end{align*}

 \begin{remark} $d$ does {\em not} restrict to $\Lambda^*(\g,K;\h)$.
 Also, $d \circ \pi^*_\g \neq \pi^*_\g \circ d$.
 \end{remark}
 We have the commutative diagram
 \begin{align*}
     \xymatrix{
     \Omega^*_{tens.}(P_\homo;\h)^G \xynor[rr]^{\tilde\Phi} \xynor[d]^{d_\omega} && \Lambda^*_\h(\g\times\h,\hat{K};\h) \xynor[d]^{d_{\tWang}} \xynor[rr]^{(\pi_\g^*)^{-1}} && \Lambda^*(\g,K;\h) \xynor[d]^{d_\Wang}\\
     \Omega^*_{tens.}(P_\homo;\h)^G \xynor[rr]^{\tilde\Phi} && \Lambda^*_\h(\g\times\h,\hat{K};\h) \xynor[rr]^{(\pi_\g^*)^{-1}} && \Lambda^*(\g,K;\h)
     }
 \end{align*}

 \begin{prop} \label{Phi-prop} The map $\Phi : \Omega^*_{tens.}(P_\homo;\h)^G \ra \Lambda^*(\g,K;\h)$, $\Phi = (\pi_\g^*)^{-1} \circ \tilde\Phi$ is a graded algebra isomorphism which commutes with the exterior covariant derivative.
 \end{prop}

 Explicitly, what does $d_\Wang$ look like?  Let $z_i = (x_i,y_i)$.  Then using \eqref{extd} and recalling that $\Hor{\tWang}(z_i) = (x_i, \Wang(x_i))$, we have
 \begin{align*}
    (\pi_\g^*(d_\Wang\varphi))(z_1,...,z_{p+1})
   &= (d_\tWang(\pi_\g^*\varphi))(z_1,...,z_{p+1}) \\
   &= (d(\pi_\g^*\varphi))(\Hor{\tWang}(z_1),...,\Hor{\tWang}(z_{p+1})) \\
%   &= \sum_{i=1}^{p+1} (-1)^{i-1} [proj_\h(\Hor{\tWang}(z_i)),(\pi_\g^*\varphi)(\Hor{\tWang}(z_1),...,\widehat{\Hor{\tWang}(z_i)},...,\Hor{\tWang}(z_{p+1}))] \nonumber\\
%   & + \sum_{i < j} (-1)^{i+j} (\pi_\g^*\varphi)([\Hor{\tWang}(z_i),\Hor{\tWang}(z_j)],\Hor{\tWang}(z_1),...,\widehat{\Hor{\tWang}(z_i)},...,\widehat{\Hor{\tWang}(z_j)},...,\Hor{\tWang}(z_{p+1}))\\
    &= \sum_{i=1}^{p+1} (-1)^{i-1} [\Wang(x_i),\varphi(x_1,...,\hat{x}_i,...,x_{p+1})]\\
    & \quad + \sum_{i < j} (-1)^{i+j} \varphi([x_i,x_j],x_1,...,\hat{x}_i,...,\hat{x}_j,...,x_{p+1}). \nonumber
 \end{align*}
 This motivates us to define on all of $\Lambda^*(\g;\h)$,
 \begin{align}
   d_\Wang \varphi = d\varphi + [\Wang,\varphi], \quad\quad \forall \varphi \in \Lambda^*(\g;\h).
   \label{dW}
 \end{align}

 \begin{remark}
 If $\rho_\Wang$ (from Lemma \ref{L-rep}) were a representation, then the formula for $d_\Wang$ would coincide with the exterior derivative defined with respect to $\rho_\Wang$.
 \end{remark}

 For a Wang map $\Wang$, we {\em cannot} define the curvature $F_\Wang$ of $\Wang$ via the formula $d_\Wang \Wang = d\Wang + [\Wang,\Wang]$ since this is not in general $\k$-semibasic.  (It is $K$-invariant, however.)  Instead, we define
 \begin{align*}
   F_\Wang := (\pi_\g^*)^{-1}(F_\tWang) \in \Lambda^2(\g;\h),
 \end{align*}
 where $\tWang(x,y) = \Wang(x)- y$.  The following are straightforward to prove.

 \begin{prop} We have the following identities:
 \begin{enumerate}
 \item $F_\Wang(x_1,x_2) = [\Wang(x_1),\Wang(x_2)] - \Wang([x_1,x_2])$, so $F_\Wang=0$ iff $\Wang$ is a Lie algebra homomorphism.
 \item $F_\Wang = d\Wang + \frac{1}{2} [\Wang,\Wang]$ and $d_\Wang F_\Wang=0$
 \item For $\varphi \in \Lambda^*(\g;\h)$, $d_\Wang\varphi = d\varphi + [\Wang,\varphi]$ and $d_\Wang^2\varphi = [F_\Wang,\varphi]$.
 \item If $\Wang : \g \ra \h$ is a Wang map, then
    \begin{itemize}
    \item $F_\Wang \in \Lambda^2(\g,K;\h)$ and $F_\Wang = \Phi(F_\omega)$, where $\omega$ is the $G$-invariant connection on $P_\homo$ corresponding to $\Wang$.
    \item $d_\Wang$ preserves $\Lambda^*(\g,K;\h)$ and $F_\Wang \in \Lambda^2(\g,K;\h)$.
    \item Let $\Lieder{x}^\Wang\varphi := \Lieder{x}\varphi + [\Wang(x),\varphi]$.  We have $\Lieder{x}^\Wang\varphi = i_x d_\Wang\varphi + d_\Wang i_x \varphi$.  If $K$ is connected, then $\varphi$ is $K$-invariant iff $\Lieder{x}^\Wang\varphi=0$ for all $x \in \k$.
    \end{itemize}
 \end{enumerate}
 \end{prop}

 \begin{example} \label{lambda-curvature} Suppose as in Lemma \ref{lambda-reductive-lemma} that $G/K$ is $\homo$-reductive with decomposition $\g = \k \oplus \s$.  Then 
 \begin{align*}
 F_{\Wang_\homo}(x_1,x_2) = -\homo_*(proj_\k([x_1,x_2])), \quad \forall x_1,x_2 \in \s,
 \end{align*}
  so in general the canonical $G$-invariant connection is not flat.
 \end{example}

 \subsection{The Hodge star and codifferential operators}

 It will be useful to adopt a different viewpoint when considering the Hodge star and codifferential operators and the Yang--Mills Lagrangian.  Given a PFB $P=P(M,H)$, let $Ad(P) = P \times_{Ad} \h$ denote the adjoint bundle which is defined as the quotient of $P \times \h$ by the equivalence relation $(p,y) \sim (ph,Ad_{h^{-1}}(y)),\forall h \in H$.  The adjoint bundle is a vector bundle over $M$ associated to $P$ with typical fibre $\h$.  Let $\Omega^*(M;Ad(P))$ denote the forms on the base manifold with values in the adjoint bundle.  Using the bundle projection $\pi : P \ra M$, we have the identification
 \begin{align}
 \Omega^*(M;Ad(P)) \cong \Omega^*_{tens.}(P;\h), \quad\quad \omega \mapsto \pi^*\omega.
 \label{adjoint-valued-forms}
 \end{align}
 Note that $(\pi^*\omega)_p : T_p P \ra Ad(P)_x$ and at the point $p$, we identify $Ad(P)_x = \{ \overline{(p,y)} \st y \in \h \}$ with $\h$ via the canonical isomorphism $\overline{(p,y)} \mapsto y$.
 The reason for adopting this new viewpoint is that: (1) the induced Hodge star operation in the Lie algebra setting is obtained more quickly, and (2) the definition of the Yang--Mills Lagrangian will be cleaner - i.e.\ it will not involve pullback by arbitrary local sections of $P \ra M$.  For an alternative way of defining these directly using objects on the principal bundle, see Bleecker \cite{Bleecker1981}.

 The space of sections of $Ad(P)$ is the Lie algebra of (right) $H$-invariant vertical vector fields on $P$.  This induces a graded commutator structure on $\Omega^*(M;Ad(P))$.  There are also corresponding definitions of exterior derivative, exterior covariant derivative, etc. (see for example \cite{AB1982} or \cite{Laquer1984}) and the identification \eqref{adjoint-valued-forms} is completely natural in that it respects all of these structures.

 A (pseudo-Riemannian) metric $\mu$ on $M$ induces a scalar product on $\Omega^*(M)$ in the usual way.  Together with an $Ad(H)$-invariant metric $m$ on $\h$, we get a well-defined scalar product $\langle \cdot, \cdot \rangle$ on $\Omega^*(M;Ad(P))$.  Moreover, using the metric $m$, we have a natural pairing
 \begin{align*}
 \Omega^k(M;Ad(P))\otimes \Omega^\ell(M;Ad(P)) \stackrel{\wedge}{\lra} \Omega^{k+\ell}(M).
 \end{align*}
 Let $n=dim(M)$ and let $\nu$ be a volume form on $M$.  The Hodge star operator $* : \Omega^k(M;Ad(P)) \ra \Omega^{n-k}(M;Ad(P))$ is uniquely defined by the condition
 \begin{align*}
 \omega \wedge *\eta = \langle \omega, \eta \rangle \nu.
 \end{align*}

 Note that the $G$-action on $P$ induces a natural $G$-action on $Ad(P)$ by $\overline{(p,y)} \mapsto \overline{(gp,y)}$, and hence on $\Omega^*(M;Ad(P))$.  If $\mu$ and $\nu$ are $G$-invariant, then $\langle \cdot , \cdot \rangle$ is $G$-invariant, and for any $g \in G$,
 \begin{align*}
 \omega \wedge (g^{-1} \cdot (* (g\cdot\eta)))
    &= g^{-1} \cdot ( g\cdot\omega \wedge *(g\cdot\eta) )
    = g^{-1} \cdot ( \langle g\cdot\omega, g\cdot\eta \rangle \nu)\\
    &= \langle \omega, \eta \rangle (g^{-1} \cdot \nu)
    = \langle \omega, \eta \rangle \nu = \omega \wedge *\eta,
 \end{align*}
 i.e.\ $*$ commutes with the $G$-action.  Thus, $*$ preserves the space of $G$-invariant forms.  On a homogeneous PFB $P_\homo(G/K,H)$, it suffices to define the Hodge star operation at $\overline{e} \in G/K$, i.e.\ on $\Lambda^*(T_{\overline{e}}(G/K);Ad(P)_{\overline{e}})$, or equivalently, on $\Lambda^*_\k(\g;\h)$.  We take this definition to be induced from the correspondence $\Phi$ (c.f.\  Prop. \ref{Phi-prop})

 We have a natural $K$-action on $\g/\k$ by $k \cdot \overline{x} = \overline{Ad_k(x)}$.
 Let $\s$ be any (not necessarily reductive) vector space complement to $\k$ in $\g$, i.e.\ $\g = \k \oplus \s$.
 Define a linear isomorphism $T : \g/\k \ra \s$ by $\overline{x} \mapsto proj_\s(x)$.  This induces a natural $K$-action on $\s$ through $k \cdot y = T(k\cdot (T^{-1}(y)))$, or
 \begin{align}
   k \cdot y = proj_\s(Ad_k(y)), \qbox{for} y \in \s \subset \g.
 \label{K-action-on-s}
 \end{align}
 Thus, these representations of $K$ on $\g/\k$ and $\s$ are isomorphic and hence the actual choice of vector space complement $\s$ will be irrelevant in the discussion to follow.  A $G$-invariant metric $\mu$ on $G/K$ is in 1-1 correspondence with a $K$-invariant scalar products on $\s$, which we will also denote by $\mu$.  Similarly, a $G$-invariant volume form $\nu$ on $G/K$ corresponds to a $K$-invariant volume form on $\s$, also denoted $\nu$.
The metric $\mu$ yields the musical isomorphism:
 \begin{align*}
   x \in \s \quad \stackrel{\flat}{\mapsto} \quad x^\flat \in \s^*=\Lambda^1_\k(\g), \quad\mbox{where } x^\flat(y) = \left\{ \begin{array}{cc}
   0 & \mbox{if } y \in \k\\
   \mu(x, y) & \mbox{if } y \in \s
   \end{array}
   \right.
 \end{align*}
 with inverse map $\sharp = \flat^{-1}$.  This isomorphism extends to the entire tensor algebra on $\s$ and restricts to the subspace of $K$-invariant objects, allowing us to raise and lower indices in the component calculations to follow.  Let
 \begin{itemize}
 \item $\{e_{\tilde{\alpha}} \}_{\tilde\alpha=1}^{dim(\k)}$ and $\{e_\alpha\}_{\alpha=1}^n$ denote bases of $\k$ and $\s$ respectively.
 \item $\{ e^\alpha \}_{\alpha=1}^n$ denote the corresponding dual basis in $\s^*$, i.e.\ $e^\alpha(e_\beta) = \delta^\alpha_\beta$.  By the identification $\s^* = \Lambda^1_\k(\g)$, we have $\Lambda^*(\s;\h) = \Lambda^*_\k(\g;\h)$.  (Note $e^\alpha \neq \mu(e_\alpha, \cdot)$ in general.  Also, for $\varphi = \varphi_\alpha e^\alpha$, we have $\varphi^\sharp = \varphi_\gamma \mu^{\gamma\alpha} e_\alpha$.)
 \item $\epsilon$ be the (completely antisymmetric, covariant) Levi-Civita permutation symbol, normalized in the chosen basis by $\epsilon_{12\cdots n}=1$.
 \item $\{ f_a \}$ denote a basis of $\h$
 \item the commutator relations on $\g$ and $\h$ be written as
    \begin{align*}
    &[e_{\tilde\alpha},e_{\tilde\beta}] =
        c_{\tilde\alpha\tilde\beta}{}^{\tilde\gamma} e_{\tilde\gamma},
    &&[e_{\tilde\alpha},e_{\beta}] =
        c_{\tilde\alpha \beta}{}^{\tilde\gamma} e_{\tilde\gamma} +
        c_{\tilde\alpha \beta}{}^{\gamma} e_\gamma,\\{}
    &[e_{\alpha},e_{\beta}] =
        c_{\alpha \beta}{}^{\tilde\gamma} e_{\tilde\gamma} +
        c_{\alpha \beta}{}^{\gamma} e_\gamma,
    &&[f_a,f_b] = r_{ab}{}^c f_c.
    \end{align*}
 \end{itemize}

 Given $\varphi = \varphi^a{}_{\beta_1\cdots \beta_k} e^{\beta_1}\otimes ... \otimes e^{\beta_k}\otimes f_a \in \Lambda^k_\k(\g;\h)$ where $\varphi^a{}_{\beta_1\cdots \beta_k} = \varphi^a{}_{[\beta_1\cdots \beta_k]}$, \eqref{dW} implies
 \begin{align}
    (d_\Wang \varphi)^c{}_{\gamma_1 \cdots \gamma_{k+1}} &= \sum_{i=1}^{k+1} (-1)^{i-1} \Wang^a{}_{\gamma_i} \varphi^b{}_{\gamma_1 \cdots \hat{\gamma_i} \cdots \gamma_{k+1}} r_{ab}{}^{c} \label{dW-components}\\
    & \quad +\sum_{1\leq i < j \leq k+1} (-1)^{i+j} \varphi^c{}_{\alpha \gamma_1 \cdots \hat{\gamma_i} \cdots \hat{\gamma_j} \cdots \gamma_{k+1}} c_{\gamma_i \gamma_j}{}^{\alpha} \nonumber
 \end{align}
 Note that only the components of $\Wang$ with respect to the basis of $\s$ are needed above.  For the Hodge star operator, we have
 \begin{align}
    (*\varphi)^a{}_{\alpha_1 \cdots \alpha_{n-k}}
    &= \frac{1}{k!} |\mu|^{1/2} \varphi^{a\beta_1\cdots \beta_k} \epsilon_{\beta_1 \cdots \beta_k \alpha_1 \cdots \alpha_{n-k}}
 \label{star-varphi}
 \end{align}
 where $|\mu| = det(\mu) \neq 0$, and the components of $\varphi$ on the right side have been raised using the metric.  If $\varphi$ is $K$-invariant, then so is $*\varphi$. 
 
 Recall that $\epsilon$ is not a tensor, so caution must be exercised when raising and lowering indices using the metric in the calculations to follow.  In particular, we have the identities
 \begin{align}
 \mu^{\alpha_1\beta_1} \cdots  \mu^{\alpha_n\beta_n} \epsilon_{\beta_1 \cdots \beta_n}&= det(\mu^{-1}) \epsilon^{\alpha_1 \cdots \alpha_n},
 \label{epsilon-det}\\
 \epsilon_{\gamma_1 \cdots \gamma_{n-k} \alpha_1 \cdots \alpha_k}
 \epsilon^{\gamma_1 \cdots \gamma_{n-k} \beta_1 \cdots \beta_k}
 &= (-1)^\mu (n-k)!k! \delta_{[\alpha_1}^{\beta_1} \cdots \delta_{\alpha_k]}^{\beta_k},
 \label{epsilon-epsilon}
 \end{align}
 where $(-1)^\mu$ is the parity corresponding to the number of negative signs in the signature of $\mu$.  These can be used to show the standard Hodge star identity:
 \begin{align*}
 **\varphi = (-1)^{k(n-k)} (-1)^\mu \varphi.
 \end{align*}

 Given a connection $\omega$ on $P_\homo$, and metric $\mu$ and volume form $\nu$ on $G/K$, the covariant codifferential $\delta_\omega$ is defined by
 \begin{align*}
     \delta_\omega \varphi = -(-1)^\mu (-1)^{n(k+1)} *d_\omega*\varphi, \quad \forall\varphi \in \Omega^k_{tens.}(P_\homo;\h),
 \end{align*}
 where $(-1)^\mu$ is the parity corresponding to the number of negative signs in the signature of $\mu$.
 If $\omega$, $\mu$ and $\nu$ are all $G$-invariant, then $\delta_\omega$ preserves $\Omega^*_{tens.}(P_\homo;\h)^G$.  Since $\Phi$ commutes with the Hodge star and exterior covariant derivative, we
define
 \begin{align}
    \delta_\Wang \varphi = -(-1)^\mu (-1)^{n(k+1)}* d_\Wang *\varphi, \quad \forall\varphi \in \Lambda^k(\g,K;\h),
    \label{codiff-W}
 \end{align}
 where $\Wang = \Phi(\omega)$.  Let us evaluate this in component form using \eqref{dW-components} and \eqref{star-varphi}.
 \begin{align*}
    &({}* d_\Wang*\varphi )^c{}_{\sigma_1 \cdots \sigma_{k-1}}
    = \frac{|\mu|^{1/2}}{(n-k+1)!} (d_\Wang *\varphi )^{c\gamma_1 \cdots \gamma_{n-k+1}} \epsilon_{\gamma_1 \cdots \gamma_{n-k+1}\sigma_1 \cdots \sigma_{k-1}}\\
    &= \frac{|\mu|^{1/2}}{(n-k+1)!} \mu^{\alpha_1 \gamma_1} \cdots \mu^{\alpha_{n-k+1} \gamma_{n-k+1}} \epsilon_{\gamma_1 \cdots \gamma_{n-k+1}\sigma_1 \cdots \sigma_{k-1}} \nonumber\\
    & \qquad \cdot \left( \sum_{i=1}^{n-k+1} (-1)^{i-1} \Wang^a{}_{\alpha_i} (*\varphi)^b{}_{\alpha_1 \cdots \hat{\alpha}_i \cdots \alpha_{n-k+1}} r_{ab}{}^{c} \right. \nonumber\\
    & \qquad\qquad \left. + \sum_{1\leq i < j \leq n-k+1} (-1)^{i+j} (*\varphi)^c{}_{\rho \alpha_1 \cdots \hat{\alpha}_i \cdots \hat{\alpha}_j \cdots \alpha_{n-k+1}} c_{\alpha_i\alpha_j}{}^{\rho}\right)  \nonumber\\
    %%%%%%%%%%%%%%%
    &= \frac{|\mu|}{k!(n-k+1)!} \mu^{\alpha_1 \gamma_1} \cdots \mu^{\alpha_{n-k+1} \gamma_{n-k+1}} \epsilon_{\gamma_1 \cdots \gamma_{n-k+1}\sigma_1 \cdots \sigma_{k-1}} \nonumber\\
    & \qquad \cdot \left( \sum_{i=1}^{n-k+1} (-1)^{i-1} \Wang^a{}_{\alpha_i} \varphi^{b\beta_1 \cdots \beta_k} \epsilon_{\beta_1 \cdots \beta_k \alpha_1 \cdots \hat{\alpha}_i \cdots \alpha_{n-k+1}} r_{ab}{}^{c} \right. \nonumber\\
    & \qquad\qquad \left. + \sum_{1\leq i < j \leq n-k+1} (-1)^{i+j} \varphi^{c\beta_1 \cdots \beta_k} \epsilon_{\beta_1 \cdots \beta_k \rho \alpha_1 \cdots \hat{\alpha}_i \cdots \hat{\alpha}_j \cdots \alpha_{n-k+1}} c_{\alpha_i\alpha_j}{}^{\rho}\right)  \nonumber\\
    %%%%%%%%%%%%%%%
    &= \frac{1}{k!(n-k+1)!} \epsilon_{\gamma_1 \cdots \gamma_{n-k+1}\sigma_1 \cdots \sigma_{k-1}} \nonumber\\
    & \qquad \cdot \left( \sum_{i=1}^{n-k+1} (-1)^{i-1} \Wang^{a\gamma_i} \varphi^b{}_{\beta_1 \cdots \beta_k} \epsilon^{\beta_1 \cdots \beta_k \gamma_1 \cdots \hat{\gamma}_i \cdots \gamma_{n-k+1}} r_{ab}{}^{c} \right. \nonumber\\
    & \quad\qquad \left. + \sum_{1\leq i < j \leq n-k+1} (-1)^{i+j} \varphi^c{}_{\beta_1 \cdots \beta_k} \epsilon^{\beta_1 \cdots \beta_k \rho \gamma_1 \cdots \hat{\gamma}_i \cdots \hat{\gamma}_j \cdots \gamma_{n-k+1}} c^{\gamma_i\gamma_j}{}_\rho \right) \nonumber
 \end{align*}
 where we have used \eqref{epsilon-det} and have raised and lowered indices on $c_{\alpha\beta}{}^\gamma$ using $\mu$. Using \eqref{epsilon-epsilon}, we have
 \begin{align*}
 &\epsilon_{\gamma_1 \cdots \gamma_{n-k+1}\sigma_1 \cdots \sigma_{k-1}} \epsilon^{\beta_1 \cdots \beta_k \gamma_1 \cdots \hat{\gamma}_i \cdots \gamma_{n-k+1}} \\%&= (-1)^{k(n-k) + n-i}  \epsilon_{\gamma_1 \cdots \hat\gamma_i \cdots \gamma_{n-k+1}\sigma_1 \cdots \sigma_{k-1}\gamma_i} \epsilon^{\gamma_1 \cdots \hat{\gamma}_i \cdots \gamma_{n-k+1} \beta_1 \cdots \beta_k}\\
 &\qquad= (-1)^{k(n-k) + n-i} (-1)^\mu (n-k)!k! \delta_{[\sigma_1}^{\beta_1} \cdots \delta_{\sigma_{k-1}}^{\beta_{k-1}} \nonumber\delta_{\gamma_i]}^{\beta_k},\\
 %%%%%%%%%%%%%%%%
 &\epsilon_{\gamma_1 \cdots \gamma_{n-k+1}\sigma_1 \cdots \sigma_{k-1}} \epsilon^{\beta_1 \cdots \beta_k \rho \gamma_1 \cdots \hat{\gamma}_i \cdots \hat{\gamma}_j \cdots \gamma_{n-k+1}}\\
% &\qquad= (-1)^{(n-k-1)(k+1) + n-j + n-i-1} \epsilon_{\gamma_1 \cdots \hat{\gamma}_i \cdots \hat{\gamma}_j \cdots \gamma_{n-k+1}\sigma_1 \cdots \sigma_{k-1}\gamma_i \gamma_j} \epsilon^{\gamma_1 \cdots \hat{\gamma}_i \cdots \hat{\gamma}_j \cdots \gamma_{n-k+1}\beta_1 \cdots \beta_k \rho }\\
  &\qquad= (-1)^{(n-k-1)(k+1) -j -i-1} (-1)^\mu (n-k-1)!(k+1)!\delta_{[\sigma_1}^{\beta_1} \cdots \delta_{\sigma_{k-1}}^{\beta_{k-1}} \delta_{\gamma_i}^{\beta_{k}} \nonumber\delta_{\gamma_j]}^{\rho},
 \end{align*}
 which we use to simplify the previous calculation.  Substituting into \eqref{codiff-W}, we obtain
 \begin{align*}
 (\delta_\Wang\varphi)^c{}_{\sigma_1 \cdots \sigma_{k-1}} &= -(-1)^\mu (-1)^{n(k+1)}(*(d_\Wang (*\varphi)))^c{}_{\sigma_1 \cdots \sigma_{k-1}}\\
%     &= - \frac{(-1)^{k^2+1}}{(n-k+1)}\sum_{i=1}^{n-k+1}  \Wang^{a\gamma_i} \varphi^b{}_{\beta_1\cdots \beta_k} r_{ab}{}^{c} \delta^{\beta_1}_{[\sigma_1} \cdots \delta^{\beta_{k-1}}_{\sigma_{k-1}} \delta^{\beta_k}_{\gamma_i]} \\
 %  & \quad + \frac{(-1)^{k+1}(k+1)}{(n-k+1)(n-k)} \sum_{1\leq i < j \leq n-k+1} \varphi^c{}_{\beta_1\cdots \beta_k} c^{\gamma_i \gamma_j}{}_{\rho} \delta^{\beta_1}_{[\sigma_1} \cdots \delta^{\beta_{k-1}}_{\sigma_{k-1}} \delta^{\beta_k}_{\gamma_i} \delta^\rho_{\gamma_j]} \nonumber\\
    &= \frac{(-1)^{k^2}}{(n-k+1)} \sum_{i=1}^{n-k+1}   \Wang^{a\tau} \varphi^b{}_{\sigma_1\cdots \sigma_{k-1} \gamma_i} r_{ab}{}^{c} \\
    & \quad+ \frac{(-1)^{(k+1)^2}(k+1)}{(n-k+1)(n-k)} \sum_{1\leq i < j \leq n-k+1}  \varphi^c{}_{[\sigma_1\cdots \sigma_{k-1} \gamma_i} c^{\gamma_i \gamma_j}{}_{\gamma_j]}
    \end{align*}
 and so
 \begin{align}
 (\delta_\Wang\varphi)^c{}_{\sigma_1 \cdots \sigma_{k-1}} = \frac{(-1)^{k+1}(k+1)}{2} \varphi^c{}_{[\sigma_1\cdots \sigma_k} c^{\sigma_k \sigma_{k+1}}{}_{\sigma_{k+1}]} - \Wang^{a\tau} \varphi^b{}_{\tau\sigma_1\cdots \sigma_{k-1}} r_{ab}{}^{c}.
 \label{codiff}
 \end{align}

 \begin{example} \label{symmetric-space}
 For any (pseudo-Riemannian) symmetric space $G/K$, there exists a canonical decomposition $\g = \k\oplus\s$ such that
 \begin{align*}
   [\k,\k] \subset \k, \quad [\k,\s] \subset \s, \quad [\s,\s] \subset \k.
 \end{align*}
 Consequently, with respect to bases $\{ e_{\tilde\alpha} \}$ on $\k$ and $\{ e_\alpha \}$ on $\s$ respectively, we have $c_{\alpha\beta}{}^\gamma = 0$.  Thus, the first term in \eqref{codiff} vanishes.  In this case,
 \begin{align*}
   (\delta_\Wang \varphi)(x_1,...,x_k) = -\sum_\tau [(\Wang|_\s)^\sharp(e^\tau),\varphi(e_\tau,x_1,...,x_k)].
 \end{align*}
 \end{example}

 \subsection{The Yang--Mills equations}

Suppose that on $M$ we have a metric $\mu$ and volume form $\nu$, and that $H$ is a compact semi-simple Lie group.  Let $m$ denote any $Ad(H)$-invariant inner product on $\h$.  (The negative of the Killing form is one such inner product.)  Given a principal bundle $P=P(M,H)$, let $\langle \cdot , \cdot \rangle$ denote the scalar product on $\Omega^*(M;Ad(P))$ induced from $\mu$ and $m$.  The Yang--Mills Lagrangian for connections $\omega$ with curvature $F_\omega \in \Omega^2(M;Ad(P))$ is
 \begin{align*}
   \lagrangian_{YM}[\omega] = \langle F_\omega, F_\omega \rangle \nu,
 \end{align*}
 and the Euler--Lagrange equations corresponding to $\lagrangian_{YM}$ are
 \begin{align*}
 \delta_\omega F_\omega = 0.
 \end{align*}

Let us consider the reduction of the Yang--Mills equations on a homogeneous PFB $P_\homo = P_\homo(G/K,H)$ using our isomorphism $\Phi$.  Given a $G$-invariant connection $\omega$ with curvature $F_\omega \in \Omega^2(G/K;Ad(P_\homo))^G$, we have a corresponding Wang map $\Wang :\g \ra \h$ and curvature $F_\Wang = F^a{}_{\alpha\beta} e^\alpha \otimes e^\beta \otimes f_a \in \Lambda^*(\g,K;\h)$.  Finally, the reduced Yang--Mills equations are $\delta_\Wang F_\Wang =0$, or in components using \eqref{codiff},
 \begin{align}
 0 = (\delta_\Wang F_\Wang)^c{}_\alpha = - \frac{3}{2} F^c{}_{[\alpha\tau} c^{\tau \sigma}{}_{\sigma]} - \Wang^{a\tau} F^b{}_{\tau\alpha} r_{ab}{}^{c} .
 \label{YM-reduced}
 \end{align}

 If $G/K$ is {\em reductive}, or more generally {\em $\homo$-reductive}, and $H$ is compact semi-simple, when does the canonical connection $\omega_\homo$ on $P_\homo(G/K,H)$ satisfy the Yang--Mills equations?  Let $\Wang_\homo$ be the corresponding Wang map (as defined in Lemma \ref{lambda-reductive-lemma}).  On the basis $\{ e_\alpha \}$ for the complement $\s$, we have as in Example \ref{lambda-curvature},
 \begin{align}
    F_{\Wang_\homo}(e_\alpha,e_\beta)
    &= -\homo_*(proj_\k([e_\alpha,e_\beta])) 
    = -\homo_{\tilde\gamma}{}^c c_{\alpha\beta}{}^{\tilde\gamma} f_c =: F^c{}_{\alpha\beta} f_c.
    \label{F-canonical}
 \end{align}
 The second term in \eqref{YM-reduced} vanishes since $\Wang_\homo|_\s=0$, so we have
 \begin{align*}
 0 &= 2(\delta_{\Wang_\homo} F_{\Wang_\homo})^c{}_\alpha
 = \left( -2F^c{}_{\alpha\tau} c^{\tau\sigma}{}_\sigma - F^c{}_{\tau\sigma} c^{\tau\sigma}{}_\alpha \right).
 \end{align*}
 Substituting in \eqref{F-canonical}, we obtain:
 
 \begin{cor} \label{cor-canonical}
 Suppose $H$ is compact semi-simple and $G/K$ is $\homo$-reductive with decomposition $\g = \k \oplus \s$, and corresponding bases $\{ e_{\tilde\alpha} \}$ on $\k$ and $\{ e_\alpha \}$ on $\s$.  The canonical connection $\omega_\homo$ on $P_\homo(G/K,H)$ is Yang--Mills iff
 \begin{align}
 0 &= \homo_{\tilde\gamma}{}^c( 2  c_{\alpha\tau}{}^{\tilde\gamma} c^{\tau\sigma}{}_\sigma + c_{\tau\sigma}{}^{\tilde\gamma} c^{\tau\sigma}{}_\alpha). \label{YM-reductive-canonical}
 \end{align}
 \end{cor}

 Clearly, if $\homo$ is trivial, then $\omega_\homo$ is Yang--Mills.
 
 \begin{example}
 From Example \ref{symmetric-space}, we see that any (pseudo-Riemannian) symmetric space $G/K$ is reductive and $[\s,\s] \subset \k$, so $c_{\alpha\beta}{}^\gamma =0$ (and hence $c^{\alpha\beta}{}_\gamma=0$) and so by Corollary \ref{cor-canonical} the canonical connection $\omega_\homo$ on $P_\homo(G/K,H)$ is Yang--Mills. This recovers the result of Harnad, Tafel \& Shnider \cite{HTS1980}.  Examples include the Stiefel bundles $V_{n,q}(F) \ra G_{n,q}(F)$, where
 \begin{align*}
    V_{n,q}(F) := U_n(F) / U_{n-q}(F) \qbox{and} G_{n,q}(F) := U_n(F) / (U_{n-q}(F) \times U_q(F))
 \end{align*}
 and $F = \R, \C$, or $\H$ and $U_n(F) = SO(n), U(n)$, or $Sp(n)$ respectively.
 \end{example}

% One can easily construct examples of homogeneous PFB over reductive, non-symmetric homogeneous spaces for which the canonical connection is Yang--Mills.

% \begin{example} Let $S$ be a Lie group with an invariant metric.  Then take any $G = K \times S$ and any homogeneous PFB $P_\homo(G/K,H) \cong P_\homo(S,H)$.  The pair $(\g,\k)$ is of course reductive with $\k$ acting trivially on $\s$.  Since $\s$ is a subalgebra of $\g$, then $c_{\alpha\beta}{}^{\tilde\gamma}=0$, so the canonical $G$-invariant connection is Yang--Mills.
% \end{example}

 The following example corresponds to a homogeneous PFB over a reductive space for which the canonical connection is {\em not} Yang--Mills.

 \begin{example} Consider the 4-dimensional Lie algebra $\g = A_{4,3}$ (c.f.\  p.988 in \cite{PSWZ1976}) with basis $\{ e_1,e_2,e_3,e_4 \}$ and commutator relations
 \begin{align*}
   [e_2,e_4] = e_1, \quad [e_3,e_4]=e_3.
 \end{align*}
 Let $\k = \langle e_1 \rangle$ and take the reductive complement $\s = \langle e_2,e_3,e_4 \rangle$.  Since the $\k$-action on $\s$ is trivial, then the metric represented by the identity matrix with respect to the given basis on $\s$ corresponds to a $K$-invariant inner product on $\s$.  (Thus, indices may be raised and lowered freely.)  Take $\h = \su(2) = \langle f_1,f_2,f_3\rangle$ and $\homo_*(e_1) = f_1$.  Evaluating  \eqref{YM-reductive-canonical} when $c=1$ and $\alpha=2$, we obtain
 \begin{align*}
%   0 &=& \sum_{\sigma,\tau=2}^4 \l( 2 c_{\alpha\tau}{}^1 c_{\tau\sigma}{}^\sigma + c_{\tau\sigma}{}^1 c_{\tau\sigma}{}^\alpha \r)
%    = 2 c_{\alpha 4}{}^1 c_{43}{}^3 + 2c_{24}{}^1 c_{24}{}^\alpha = -2 c_{\alpha 4}{}^1 + 2c_{24}{}^\alpha\\
   0 &= \sum_{\sigma,\tau=2}^4 \l( 2 c_{2\tau}{}^1 c_{\tau\sigma}{}^\sigma + c_{\tau\sigma}{}^1 c_{\tau\sigma}{}^2 \r)
    = 2 c_{24}{}^1 c_{43}{}^3 = -2,
 \end{align*}
 a contradiction.  Thus, the canonical connection on $P_\homo(G/K,H)$ is not Yang--Mills.
 \end{example}

 \section{Non-reductive pseudo-Riemannian homogeneous spaces of dimension 4}
 \label{non-reductive}

 As mentioned in the introduction, the only systematic attempt so far to explicitly classify non-reductive pseudo-Riemannian homogeneous spaces in low dimensions has been recent work by Fels \& Renner \cite{FelsRenner2006}.
All such spaces are necessarily: (1) non-Riemannian, (2) of dimension 4 or higher (provided $K$ is connected).
In this section, we investigate the homogeneous PFB with compact semi-simple structure group over the 4-dimensional non-reductive examples of Fels \& Renner which admit invariant connections.  Many of these are necessarily trivial bundles.  We moreover classify all Wang maps associated with these homogeneous PFB.  With the reduced Yang--Mills equations $\delta_\Wang F_\Wang=0$ in hand from the correspondence established in Section \ref{gauge-theory}, we determine the invariant Yang--Mills connections in these cases.

 \subsection{The Fels--Renner classification}
 \label{Fels-Renner-subsection}

In dimension 4, Fels \& Renner found 8 classes of examples, labelled A1-A5 and B1-B3.  Invariant metrics of signature (1,3) are admitted by  A1-A5 while those of signature (2,2) are admitted by A1-A3 \& B1-B3.  These are displayed in the Tables \ref{FelsRenner-classification1} and \ref{FelsRenner-classification2}.  For each model we specify: (1) a (non-reductive) complement $\s$ to the isotropy subalgebra $\k$, (2) the matrices corresponding to the representation $\rho_\s : \k \ra \gl(\s)$ induced by the $K$-action in \eqref{K-action-on-s}, and (3) the $\k$-invariant metrics on $\s$.  The algebras labelled by $A^a_{x,y}$ are solvable algebras appearing on pg.\ 988 in \cite{PSWZ1976}.  We have chosen bases adapted to $\k$.  Our bases $\{ e_i \}$ in terms of the Fels--Renner bases (labelled here as $\{ \FR{i} \}$) are:
 \begin{align*}
 \begin{array}{lcl}
    A1:\quad (e_1,...,e_5)&=&(\FR{1}, \FR{2}, \FR{4}, \FR{5}, \FR{3}+\FR{4})\\
    A2:\quad (e_1,...,e_5)&=&(\FR{1}, \FR{2}, \FR{3}, \FR{5}, \FR{4})\\
    A3:\quad (e_1,...,e_5)&=&(\FR{1}, \FR{2}, \FR{4}, \FR{5}, \FR{3})\\
    A4:\quad (e_1,...,e_5,e_6)&=&(\FR{1}, \FR{2}, \FR{6}, \FR{4}, \FR{3}+\FR{6}, \FR{5})\\
    A5:\quad (e_1,...,e_5,e_6,e_7)&=&(\FR{7}, \FR{2}, \FR{4}, \FR{6}, \FR{1}+\FR{7}, \FR{3}-\FR{4}, \FR{5})\\
    B1:\quad (e_1,...,e_5)&=&(\FR{1}, \FR{2}, \FR{4}, \FR{5}, \FR{3})\\
    B2:\quad (e_1,...,e_5,e_6)&=&(\FR{1}, \FR{2}, \FR{4}, \FR{6}, \FR{3}-\FR{6}, \FR{5})\\
    B3:\quad (e_1,...,e_5,e_6,e_7)&=&(\FR{1}, \FR{2}, \FR{4}, \FR{5}, \FR{3}, \FR{5}+\FR{6})
 \end{array}
 \end{align*}
 Note that for B1 on pg.\ 285 in \cite{FelsRenner2006}: $[e_1,e_3]=-2e_2$ should read $[e_1,e_3]=-2e_3$. 

 \subsection{Homogeneous PFB and invariant connections}

 We suppose that: (1) $K$ is connected, and (2) $H$ is compact semi-simple.  In this section we will obtain some general classification results regarding the homogeneous PFB $P_\homo(G/K,H)$ which admit a $G$-invariant connection.  Let $\homo : K \ra H$ be a Lie group homomorphism. Suppose that $\Wang : \g \ra \h$ is a Wang map which extends $\homo_* : \k \ra \h$.  Since $K$ is connected, $K$-invariance is equivalent to $\k$-invariance, i.e.
 \begin{align}
   [\homo_*(x), \Wang(v)] = \Wang([x,v]), \qquad \forall x \in \k,\quad \forall v \in \g.
   \label{L-inv}
 \end{align}
 Note that for $x \in \k$ and $v \in \g$,
 \begin{itemize}
 \item if $v \in \ker(ad_x)$, then $\Wang(v) \in \ker(ad_{\homo_*(x)})$
 \item if $v \in \im(ad_x)$, then $\Wang(v) \in \im(ad_{\homo_*(x)})$
 \item if $v \in \im(ad_x) \cap \ker(ad_x)$ then $\Wang(v) \in \im(ad_{\homo_*(x)})\cap \ker(ad_{\homo_*(x)})$.  Moreover, if $\homo_*(x),\Wang(v) \neq 0$, then $\Wang(v)$ is a generalized eigenvector (corresponding to the generalized eigenvalue 0) of the map $ad_{\homo_*(x)}$.
 \end{itemize}
 However, if $\h$ is semi-simple, then any representation of $\h$ is via diagonalizable matrices.  Thus, in the last case above, $ad_{\homo_*(x)}$ is diagonalizable.  Thus, if $v \in \im(ad_x) \cap \ker(ad_x)$, then either $\homo_*(x)=0$ or $\Wang(v)=0$.  In the former case, we have from the Wang condition that $\Wang=0$ on $\im(ad_x)$.

 \begin{lemma} \label{lem1}
 Suppose $\h$ is semi-simple and $\Wang : \g \ra \h$ is a Wang Map which extends $\homo_* : \k \ra \h$.  If $x \in \k$ and $v \in \im(ad_x) \cap \ker(ad_x)$, then either: (1) $\homo_*(x)=0$, or (2) $\Wang(v)=0$.  If $\homo_*(x)=0$, then $\Wang=0$ on $\im(ad_x)$.
 \end{lemma}

 Since $H$ is compact semi-simple, then the Killing form is nondegenerate, negative-definite and invariant under automorphisms, so $ad_v \in \so(dim\, \h)$.

 \begin{lemma} \label{lem2} Suppose $\h$ is compact semi-simple, then for any $y \in \h$, the map $ad_y \in End(\h)$ has only purely imaginary eigenvalues.
 \end{lemma}

 Applying these two facts in each of the 8 cases, we have a classification of all possible Wang maps.  If $\Wang=0$ on $\k$, then $\homo_*$ is trivial and the image of $\homo$ is discrete.  But $K$ is connected, so $\homo$ must be trivial, hence $P_\homo(G/K,H) \ra G/K$ is trivial.

% \newpage
 \begin{landscape}
 \begin{table}[h]
 \centering
 \hspace{-0.5in}
 \begin{footnotesize}
 $\begin{array}{|@{}l@{}|@{}l@{}|@{}l@{}|@{}c@{}|} \hline
  \multicolumn{1}{|c|}{\mbox{Non-reductive pair } (\g,\k)} & \multicolumn{1}{c|}{\mbox{Commutation relations}} & \multicolumn{1}{c|}{\mbox{Representation } \rho_\s : \k \ra \gl(\s)} & \k\mbox{-invariant metric on } \s\\ \hline\hline
 \begin{array}{l}
 A1:    \begin{array}{l} \g=\sl(2,\R) \times \solv(2) \\
        \k \cong \R
        \end{array}
 \end{array}
        & \begin{array}{l}
        \begin{array}{c|ccccc}
        & e_1 & e_2 & e_3 & e_4 & e_5\\ \hline
        e_1 & \cdot & 2e_2 & \cdot & \cdot & 2(e_3-e_5)\\
        e_2 &  & \cdot & \cdot & \cdot & e_1\\
        e_3 &  &  & \cdot & e_3 & \cdot\\
        e_4 &  &  &  & \cdot & -e_3\\
        e_5 &  &  &  & & \cdot
        \end{array}\\
        \quad\quad\quad\quad(\k: e_5; \quad \s: e_1,...,e_4)
        \end{array}
    & \begin{array}{l}
     \rho_\s(e_5) = \l( \begin{array}{cccc}
    0 & -1 & 0 & 0\\
    0 & 0 & 0 & 0\\
    -2 & 0 & 0 & 1\\
    0 & 0 & 0 & 0
    \end{array} \r)
    \end{array}
    & \begin{array}{c} \l( \begin{array}{cccc}
    -2a & 0 & 0 & a\\
    0 & b & a & c\\
    0 & a & 0 & 0\\
    a & c & 0 & d\\
    \end{array}
    \r)\\
    \\
    det = a^3(a+2d)
    \end{array}
    
    \\ \hline
 \begin{array}{l}
 A2: \begin{array}{l} \g = A^\alpha_{5,30}, \quad \alpha \in \R \\
     \k \cong \R
     \end{array}
 \end{array}
        & \begin{array}{l}
        \begin{array}{c|ccccc}
        & e_1 & e_2 & e_3 & e_4 & e_5\\ \hline
        e_1 & \cdot & \cdot & \cdot & (\alpha+1)e_1 & \cdot\\
        e_2 &  & \cdot & \cdot & \alpha e_2 & e_1 \\
        e_3 &  &  & \cdot & (\alpha-1)e_3 & e_2 \\
        e_4 &  &  &  & \cdot & -e_5\\
        e_5 &  &  &  &  & \cdot
        \end{array}\\
        \quad\quad\quad\quad(\k: e_5; \quad \s: e_1,...,e_4)
        \end{array}
    & \begin{array}{l}
    \rho_\s(e_5) = \l( \begin{array}{cccc}
    0 & -1 & 0 & 0\\
    0 & 0 & -1 & 0\\
    0 & 0 & 0 & 0\\
    0 & 0 & 0 & 0
    \end{array} \r)
    \end{array}
    & \begin{array}{c}
    \l( \begin{array}{cccc}
    0 & 0 & -a & 0\\
    0 & a & 0 & 0\\
    -a & 0 & b & c\\
    0 & 0 & c & d
    \end{array} \r)\\ \\
    det = -a^3 d
    \end{array}
    \\ \hline
 \begin{array}{l}
 A3: \begin{array}{l}
     \g = \l\{ \begin{array}{ll}
        A_{5,37} \quad \mbox{if } \epsilon=1\\
        A_{5,36} \quad \mbox{if } \epsilon=-1
        \end{array} \r.\\
     \k \cong \R
     \end{array}
 \end{array}
        & \begin{array}{l}
        \begin{array}{c|ccccc}
        & e_1 & e_2 & e_3 & e_4 & e_5\\ \hline
        e_1 & \cdot & \cdot & 2e_1 & \cdot & \cdot\\
        e_2 &  & \cdot & e_2 & -\epsilon e_5 & e_1\\
        e_3 &  &  & \cdot & \cdot & -e_5\\
        e_4 &  &  &  & \cdot & -e_2\\
        e_5 &  &  &  &  & \cdot
        \end{array}\\
        \quad\quad\quad\quad(\k: e_5; \quad \s: e_1,...,e_4)
        \end{array}
    & \begin{array}{l}
    \rho_\s(e_5) = \l( \begin{array}{cccc}
    0 & -1 & 0 & 0\\
    0 & 0 & 0 & 1\\
    0 & 0 & 0 & 0\\
    0 & 0 & 0 & 0
    \end{array} \r)
    \end{array}
    & \begin{array}{c}
    \l( \begin{array}{cccc}
    0 & 0 & 0 & a\\
    0 & a & 0 & 0\\
    0 & 0 & b & c\\
    a & 0 & c & d
    \end{array} \r) \\\\
    det = -a^3 b
    \end{array} \\ \hline
 \begin{array}{l}
 A4: \begin{array}{l}
     \g = \n(3) \rtimes \sl(2,\R)\\
     \k \cong \R^2
     \end{array} \\
 (\n(3) = \mbox{Heisenberg algebra})
 \end{array}
  & \begin{array}{l}
        \begin{array}{c|cccccc}
        & e_1 & e_2 & e_3 & e_4 & e_5 & e_6 \\ \hline
        e_1 & \cdot & 2e_2 & \cdot & e_4 & 2(e_3-e_5) & -e_6 \\
        e_2 &  & \cdot & \cdot & \cdot & e_1 & e_4\\
        e_3 &  &  & \cdot & \cdot & \cdot & \cdot\\
        e_4 &  &  &  & \cdot & -e_6 & e_3\\
        e_5 &  &  &  &  & \cdot & \cdot\\
        e_6 &  &  &  &  &  & \cdot \\
        \end{array}\\
        \quad\quad\quad\quad(\k: e_5,e_6; \quad \s: e_1,...,e_4)
        \end{array}
    & \begin{array}{l}
     \rho_\s(e_5) = \l( \begin{array}{cccc}
    0 & -1 & 0 & 0\\
    0 & 0 & 0 & 0\\
    -2 & 0 & 0 & 0\\
    0 & 0 & 0 & 0
    \end{array} \r), \\ \hline
    \rho_\s(e_6) = \l( \begin{array}{cccc}
    0 & 0 & 0 & 0\\
    0 & 0 & 0 & 0\\
    0 & 0 & 0 & -1\\
    0 & -1 & 0 & 0
    \end{array} \r)
    \end{array}
    & \begin{array}{c}
    \l( \begin{array}{cccc}
    2a & 0 & 0 & 0\\
    0 & b & -a & 0\\
    0 & -a & 0 & 0\\
    0 & 0 & 0 & a
    \end{array} \r) \\ \\
    det = -2a^4
    \end{array} \\ \hline
 \end{array}$
 \end{footnotesize}
 \caption{Fels--Renner classification of 4-dim.\ non-reductive pseudo-Riemannian spaces $G/K$: Classes A1-A4}
 \label{FelsRenner-classification1}
 \end{table}
 \end{landscape}

% \newpage
 \begin{landscape}
 \begin{table}[h]
 \centering
 \begin{footnotesize}
 $\begin{array}{|@{}l@{}|@{}l@{}|@{}l@{}|@{}c@{}|} \hline
  \multicolumn{1}{|c|}{\mbox{Non-reductive pair } (\g,\k)} & \multicolumn{1}{c|}{\mbox{Commutation relations}} & \multicolumn{1}{c|}{\mbox{Representation } \rho_\s : \k \ra \gl(\s)} & \k\mbox{-invariant metric on } \s\\ \hline\hline
 \begin{array}{l}
 A5: \begin{array}{l}
     \g = A^1_{4,9} \rtimes \sl(2,\R)\\
     \k \cong \mbox{Bianchi V}
     \end{array}
 \end{array}
     & \begin{array}{l}
        \begin{array}{c|ccccccc}
        & e_1 & e_2 & e_3 & e_4 & e_5 & e_6 & e_7\\ \hline
        e_1 & \cdot & \cdot & -2e_3 & -e_4 & \cdot & 2e_3 & -e_7 \\
        e_2 &  & \cdot & \cdot & \cdot & -2e_2 & -e_1+e_5 & e_4\\
        e_3 &  &  & \cdot & \cdot & 2e_3 & \cdot & \cdot\\
        e_4 &  &  &  & \cdot & \cdot & -e_7 & -e_3\\
        e_5 &  &  &  &  & \cdot & -2e_6 & -2e_7\\
        e_6 &  &  &  &  &  & \cdot & \cdot\\
        e_7 &  &  &  &  &  &  & \cdot\\
        \end{array}\\
        \quad\quad\quad\quad(\k: e_5,e_6,e_7;\quad \s: e_1,...,e_4)
        \end{array}
    & \begin{array}{l}
    \rho_\s(e_5) = \l( \begin{array}{cccc}
    0 & 0 & 0 & 0\\
    0 & 2 & 0 & 0\\
    0 & 0 & -2 & 0\\
    0 & 0 & 0 & 0
    \end{array} \r), \\ \hline
    \rho_\s(e_6) = \l( \begin{array}{cccc}
    0 & 1 & 0 & 0\\
    0 & 0 & 0 & 0\\
    -2 & 0 & 0 & 0\\
    0 & 0 & 0 & 0
    \end{array} \r),\\ \hline
    \rho_\s(e_7) = \l( \begin{array}{cccc}
    0 & 0 & 0 & 0\\
    0 & 0 & 0 & 0\\
    0 & 0 & 0 & 1\\
    0 & -1 & 0 & 0
    \end{array} \r)
    \end{array}
    & \begin{array}{c}
    \l( \begin{array}{cccc}
    2a & 0 & 0 & 0\\
    0 & 0 & a & 0\\
    0 & a & 0 & 0\\
    0 & 0 & 0 & a
    \end{array} \r)\\ \\
    det = -2a^4
    \end{array} \\ \hline
 \begin{array}{l}
 B1: \begin{array}{l}
     \g = \sa(2,\R) \\
     \k \cong \R
     \end{array}
 \end{array}
 & \begin{array}{l}
        \begin{array}{c|ccccc}
        & e_1 & e_2 & e_3 & e_4 & e_5 \\ \hline
        e_1 & \cdot & 2e_2 & e_3 & -e_4 & -2e_5\\
        e_2 &  & \cdot & \cdot & e_3 & e_1\\
        e_3 &  &  & \cdot & \cdot & -e_4\\
        e_4 &  &  &  & \cdot & \cdot\\
        e_5 &  &  &  &  & \cdot
        \end{array}\\
        \quad\quad\quad\quad(\k: e_5; \quad \s: e_1,...,e_4)
        \end{array}
    & \begin{array}{l}
    \rho_\s(e_5) = \l( \begin{array}{cccc}
    0 & -1 & 0 & 0\\
    0 & 0 & 0 & 0\\
    0 & 0 & 0 & 0\\
    0 & 0 & 1 & 0
    \end{array} \r)
    \end{array}
    & \begin{array}{c}
    \l( \begin{array}{cccc}
        0 & 0 & a & 0\\
        0 & b & c & a\\
        a & c & d & 0\\
        0 & a & 0 & 0
        \end{array} \r) \\ \\
    det = a^4
    \end{array}
    \\ \hline
 \begin{array}{l}
 B2: \begin{array}{l}
     \g = \n(3) \rtimes \sl(2,\R)\\
     \k \cong \R^2
     \end{array} \\
 (\n(3) = \mbox{Heisenberg algebra})
 \end{array}
      & \begin{array}{l}
        \begin{array}{c|cccccc}
        & e_1 & e_2 & e_3 & e_4 & e_5 & e_6 \\ \hline
        e_1 & \cdot & 2e_2 & e_3 & \cdot & -2e_4-2e_5 & -e_6\\
        e_2 &  & \cdot & \cdot & \cdot & e_1 & e_3\\
        e_3 &  &  & \cdot & \cdot & -e_6 & e_4\\
        e_4 &  &  &  & \cdot & \cdot & \cdot\\
        e_5 &  &  &  &  & \cdot & \cdot\\
        e_6 &  &  &  &  &  & \cdot \\
        \end{array} \\
        \quad\quad\quad\quad(\k: e_5,e_6; \quad \s: e_1,...,e_4)
        \end{array}
    & \begin{array}{l}
    \rho_\s(e_5) = \l( \begin{array}{cccc}
    0 & -1 & 0 & 0\\
    0 & 0 & 0 & 0\\
    0 & 0 & 0 & 0\\
    2 & 0 & 0 & 0
    \end{array} \r), \\ \hline
   \rho_\s(e_6) = \l( \begin{array}{cccc}
    0 & 0 & 0 & 0\\
    0 & 0 & 0 & 0\\
    0 & -1 & 0 & 0\\
    0 & 0 & -1 & 0
    \end{array} \r)
    \end{array}
    & \begin{array}{c}
    \l( \begin{array}{cccc}
        -2a & 0 & 0 & 0\\
        0 & b & 0 & -a\\
        0 & 0 & a & 0\\
        0 & -a & 0 & 0
        \end{array} \r)\\ \\
    det = 2a^4
    \end{array}
    \\ \hline
 \begin{array}{l}
 B3: \begin{array}{l}
     \g = \sa(2,\R) \times \R \\
     \k \cong \R^2
     \end{array}
 \end{array}
     & \begin{array}{l}
        \begin{array}{c|cccccc}
        & e_1 & e_2 & e_3 & e_4 & e_5 & e_6 \\ \hline
        e_1 & \cdot & 2e_2 & e_3 & -e_4 & -2e_5 & -e_4\\
        e_2 &  & \cdot & \cdot & e_3 & e_1 & e_3\\
        e_3 &  &  & \cdot & \cdot & -e_4 & \cdot\\
        e_4 &  &  &  & \cdot & \cdot & \cdot\\
        e_5 &  &  &  &  & \cdot & \cdot\\
        e_6 &  &  &  &  &  & \cdot\\
        \end{array}\\
        \quad\quad\quad\quad(\k: e_5,e_6; \quad \s: e_1,...,e_4)
        \end{array}
    & \begin{array}{l}
    \rho_\s(e_5) = \l( \begin{array}{cccc}
    0 & -1 & 0 & 0\\
    0 & 0 & 0 & 0\\
    0 & 0 & 0 & 0\\
    0 & 0 & 1 & 0
    \end{array} \r), \\ \hline
    \rho_\s(e_6) =
    \l( \begin{array}{cccc}
    0 & 0 & 0 & 0\\
    0 & 0 & 0 & 0\\
    0 & -1 & 0 & 0\\
    1 & 0 & 0 & 0
    \end{array} \r)
    \end{array}
    & \begin{array}{c}
    \l(\begin{array}{cccc}
        0 & 0 & a & 0\\
        0 & b & 0 & a\\
        a & 0 & 0 & 0\\
        0 & a & 0 & 0
    \end{array} \r) \\ \\
    det = a^4
    \end{array} \\ \hline
 \end{array}$
 \end{footnotesize}
 \caption{Fels--Renner classification of 4-dim.\ non-reductive pseudo-Riemannian spaces $G/K$: Classes A5, B1-B3}
 \label{FelsRenner-classification2}
 \end{table}
 \end{landscape}

  \begin{table}[h]
  \begin{footnotesize}
 $\begin{array}{|c|l|c|c|c|c|} \hline
 (\g,\k) & \multicolumn{1}{|c|}{\mbox{Kernels and images of $ad(\k)$}} & \begin{array}{c} \mbox{Wang maps}\,\, \Wang : \g \ra \h \\
        (\h \mbox{ compact semi-simple})
        \end{array} & \begin{array}{c} \mbox{Trivial}\\ \mbox{PFB $P_\homo$?} \end{array} \\ \hline\hline
 A1 & \begin{array}{l@{}l@{}l}\ker(ad_{e_5}) &=& \langle e_3,e_5 \rangle,\\ \im(ad_{e_5}) &=& \langle e_1,e_3,e_5\rangle \end{array}
    & \Wang=0 \mbox{ on } \langle e_1,e_3,e_5 \rangle
    & YES
    \\ \hline
 A2 & \begin{array}{l@{}l@{}l}\ker(ad_{e_5}) &=& \langle e_1,e_5 \rangle,\\ \im(ad_{e_5}) &=& \langle e_1,e_2,e_5\rangle \end{array}
    & \Wang=0 \mbox{ on } \langle e_1,e_2,e_5 \rangle
    & YES
    \\ \hline
 A3 & \begin{array}{l@{}l@{}l}\ker(ad_{e_5}) &=& \langle e_1,e_5 \rangle,\\ \im(ad_{e_5}) &=& \langle e_1,e_2,e_5\rangle \end{array}
    & \Wang=0 \mbox{ on } \langle e_1,e_2,e_5 \rangle
    & YES
    \\ \hline
 A4 & \begin{array}{l@{}l@{}l}
    \ker(ad_{e_5}) &=& \langle e_3,e_5,e_6 \rangle,\\ \im(ad_{e_5}) &=& \langle e_1,e_3-e_5,e_6\rangle \\
    \ker(ad_{e_6}) &=& \langle e_3,e_5,e_6 \rangle,\\ \im(ad_{e_6}) &=& \langle e_3,e_4,e_6\rangle
    \end{array}
    & \Wang=0 \mbox{ on } \langle e_1,e_3,e_4,e_5,e_6 \rangle
    & YES
    \\ \hline
 A5 & \begin{array}{l@{}l@{}l}
    \ker(ad_{e_5}) &=& \langle e_1,e_4,e_5 \rangle,\\ \im(ad_{e_5}) &=& \langle e_2,e_3,e_6,e_7\rangle \\
    \ker(ad_{e_6}) &=& \langle e_3,e_6,e_7 \rangle,\\ \im(ad_{e_6}) &=& \langle e_1-e_5,e_3,e_6,e_7\rangle \\
    \ker(ad_{e_7}) &=& \langle 2e_1-e_5,e_3,e_6,e_7 \rangle,\\ \im(ad_{e_7}) &=& \langle e_3,e_4,e_7\rangle
    \end{array}
    & \begin{array}{c} \begin{array}{c}\Wang=0 \mbox{ on }\\
     \langle e_1-e_5,e_2,e_3,e_4,e_6,e_7 \rangle
    \end{array}\\ \hline \\
    \Wang=0
    \end{array}
    & \begin{array}{c} NO \\ \hline YES \end{array}
    \\ \hline
 B1 & \begin{array}{l@{}l@{}l}\ker(ad_{e_5}) &=& \langle e_4,e_5 \rangle,\\ \im(ad_{e_5}) &=& \langle e_1,e_4,e_5\rangle \end{array}
    & \Wang=0 \mbox{ on } \langle e_1,e_4,e_5 \rangle
    & YES
    \\ \hline
 B2 & \begin{array}{l@{}l@{}l}
    \ker(ad_{e_5}) &=& \langle e_4,e_5,e_6 \rangle,\\ \im(ad_{e_5}) &=& \langle e_1,e_4+e_5,e_6\rangle \\
    \ker(ad_{e_6}) &=& \langle e_4,e_5,e_6 \rangle,\\ \im(ad_{e_6}) &=& \langle e_3,e_4,e_6\rangle
    \end{array}
    & \Wang=0 \mbox{ on } \langle e_1,e_3,e_4,e_5,e_6 \rangle
    & YES
    \\ \hline
 B3 & \begin{array}{l@{}l@{}l}
    \ker(ad_{e_5}) &=& \langle e_4,e_5,e_6 \rangle,\\ \im(ad_{e_5}) &=& \langle e_1,e_4,e_5\rangle \\
    \ker(ad_{e_6}) &=& \langle e_3,e_4,e_5,e_6 \rangle,\\ \im(ad_{e_6}) &=& \langle e_3,e_4 \rangle
    \end{array}
    & \begin{array}{c} \begin{array}{c}\Wang=0 \mbox{ on } \langle e_1,e_3,e_4,e_5 \rangle \\{}
        \Wang(e_2) \in \ker(ad_{\Wang(e_6)})
      \end{array}\\ \hline
      \begin{array}{c} \Wang=0 \mbox{ on } \\ \langle e_1,e_3,e_4,e_5,e_6 \rangle \end{array} \end{array}
    & \begin{array}{c} NO \\ \hline YES \end{array} \\ \hline
 \end{array}$
 \end{footnotesize}
 \caption{Invariant connections on homogeneous PFB over 4-dim.\ non-reductive pseudo-Riemannian homogeneous spaces}
 \label{nonred-homPFB}
 \end{table}

 \begin{itemize}
 \item A1-A3, B1 cases: Straightforward application of Lemma \ref{lem1}.
 
 \item A4, B2 cases: For the A4 case, $e_6 \in \ker(ad_{e_6}) \cap \im(ad_{e_6})$, so by Lemma \ref{lem1}, $\Wang(e_6)=0$ and $\Wang=0$ on $\im(ad_{e_6})$, so $\Wang(e_3)=\Wang(e_4)=0$.  We also have $e_3-e_5 \in \ker(ad_{e_5}) \cap \im(ad_{e_5})$, so $\Wang(e_3-e_5)=0$ or $\Wang(e_5)=0$.  Since $\Wang(e_3)=0$, we must have $\Wang(e_5)=0$ and hence $\Wang=0$ on $\im(ad_{e_5})$ so $\Wang(e_1)$.  Thus, $\Wang=0$ on $\k$ and the bundle is trivial.  The B2 case is similar.

 \item A5 case: We have $e_6 \in \ker(ad_{e_6}) \cap \im(ad_{e_6})$ and $e_7 \in \ker(ad_{e_7}) \cap \im(ad_{e_7})$
  so by Lemma \ref{lem1}, $\Wang(e_6)=\Wang(e_7)=0$.   Thus, $\Wang=0$ on $\im(ad_{e_6})$ and $\im(ad_{e_7})$, so $\Wang=0$ on $\langle e_1-e_5,e_3,e_4,e_6,e_7 \rangle$.  The Wang condition \eqref{L-inv} reduces to one final constraint: $[\Wang(e_5),\Wang(e_2)] = \Wang([e_5,e_2]) = 2\Wang(e_2)$.  If $\Wang(e_2)\neq 0$, then it is an eigenvector of $ad_{\Wang(e_5)}$ with eigenvalue 2.  But by Lemma \ref{lem2}, this is impossible if $\Wang(e_5)\neq 0$ since all eigenvalues are necessarily purely imaginary.  Thus, $\Wang(e_2)=0$ or $\Wang(e_5)=0$.  In the latter case, we must have $\Wang=0$ on $\im(ad_{e_5})$, so $\Wang(e_2)=0$ as well.

 \item $B3$ case: We have $e_5 \in \ker(ad_{e_5}) \cap \im(ad_{e_5})$, so by Lemma \ref{lem1}, $\Wang(e_5)=0$.  Thus, $\Wang=0$ on $\im(ad_{e_5})$, so $\Wang(e_1)=\Wang(e_4)=0$.  Since $[e_2,e_6]=e_3$ and $[e_3,e_6]=0$, then
 $[\Wang(e_6),\Wang(e_2)] = -\Wang(e_3)$, $[\Wang(e_6),\Wang(e_3)] = 0$.  Since $\Wang(e_2)$ cannot be a generalized eigenvector of $ad_{W(e_6)}$, we have $\Wang(e_3)=0$ or $\Wang(e_6)=0$.  In the latter case, $\Wang=0$ on $\im(ad_{e_6})$, so $\Wang(e_3)=0$.  We have the remaining constraint $\Wang(e_2) \in \ker(ad_{\Wang(e_6)})$.
 \end{itemize}

 We have proved the first assertion of the following classification theorem.

 \begin{theorem} \label{classification-thm}
 Suppose $H$ compact semi-simple and $P_\homo = P_\homo(G/K,H)$ is a homogeneous PFB over a 4-dim.\ non-reductive pseudo-Riemannian homogeneous space $G/K$ (with $K$ connected).  Then:
 \begin{enumerate}
 \item There exists a nontrivial bundle $P_\homo \ra G/K$ admitting at least one $G$-invariant connection iff $G/K$ is of class A5 or B3.
% \item In either the A5 or B3 case, there exists an ideal $\k_0 \subset \k$ such that $\k_0 \subset \ker(\homo_*)$ for any homomorphism $\homo : K \ra H$ for which $P_\homo \ra G/K$ admits a $G$-invariant connection.  Moreover, $(\g,\k)$ is $\k_0$-reductive.
\item In the A5 or B3 case: if $P_\homo \ra G/K$ admits a $G$-invariant connection,
\begin{enumerate}
 \item $G/K$ is $\homo$-reductive.
   \item A5: there exists a {\em unique} $G$-invariant connection, namely the canonical connection corresponding to $\homo$-reductivity.  This connection is flat.
   \item B3: the space of $G$-invariant connections is an $r$-dim.\ vector space, where $r$ is the dimension of the centralizer $C_\h(\im(\homo_*))$.  These connections are generally not flat.
   \end{enumerate}
 \end{enumerate}
 \end{theorem}

 \begin{proof} We prove assertion 2(a) in the case that $\homo$ is nontrivial. The existence of a $G$-invariant connection on $P_\homo \ra G/K$ implies the existence of a Wang map $\Wang$ which must satisfy the conditions in Table \ref{nonred-homPFB}.  Recall also that $\Wang=\homo_*$ on $\k$, so $\ker(\homo_*) = \ker(\Wang) \cap \k$.  We need to exhibit a subspace $\s \subset \g$ such that $\g=\k\oplus\s$ and $[\k,\s] \subset \ker(\homo_*) \oplus \s$.  We have:
 \begin{itemize}
 \item A5 case: $\k=\langle e_5,e_6,e_7 \rangle$, $\ker(\homo_*) = \langle e_6, e_7 \rangle$.  Choose $\s = \langle e_1 - e_5, e_2, e_3, e_4 \rangle$.
 \item B3 case: $\k=\langle e_5,e_6 \rangle$, $\ker(\homo_*) = \langle e_5 \rangle$.  Choose $\s = \langle e_1,e_2,e_3,e_4 \rangle$.
 \end{itemize}
 Using the above decompositions and the classifications in Table \ref{nonred-homPFB}, assertions 2(b) and 2(c) follows immediately.  Flatness is evaluated using \eqref{F-canonical}.
 \end{proof}

 \subsection{Yang--Mills connections}

We will focus mainly on the nontrivial examples A5 and B3.  Suppose that for a fixed Wang map $\Wang$, $\im(\Wang)$ is an abelian subalgebra of $\h$.  This applies to the cases A4, A5, B2 \& B3.  In these cases: (1) the curvature of $\Wang$ is simply $F_\Wang(x_1,x_2) = -\Wang([x_1,x_2]),\, \forall x_1,x_2 \in \s$ and so the second term in \eqref{YM-reduced} vanishes, (2) $\s$ has been chosen to be a subalgebra of $\g$, so $c_{\alpha\tau}{}^{\tilde\rho}=0$, and so:
 \begin{align*}
 0 &= (\delta_\Wang F_\Wang)^c{}_\alpha %= -\frac{3}{2} F^c{}_{[\alpha\tau} c^{\tau \sigma}{}_{\sigma]}
%    = \frac{3}{2} \l( \Wang^c{}_\rho c_{[\alpha\tau}{}^\rho c^{\tau \sigma}{}_{\sigma]} + \Wang^c{}_{\tilde\rho} c_{[\alpha\tau}{}^{\tilde\rho} c^{\tau \sigma}{}_{\sigma]} \r)\\
 = \frac{3}{2} \Wang^c{}_\rho c_{[\alpha\tau}{}^\rho c^{\tau \sigma}{}_{\sigma]} 
 = \Wang^c{}_\rho c_{\alpha\tau}{}^\rho c^{\tau \sigma}{}_{\sigma} + \frac{1}{2} \Wang^c{}_\rho c_{\sigma\tau}{}^\rho c^{\sigma\tau}{}_{\alpha}.
 \end{align*}

 \begin{table}[h]
 \begin{center}
 \renewcommand{\arraystretch}{1.25}
 $\begin{array}{|c|c|c|c|} \hline
  & c_{\alpha\beta}{}^\gamma
              & c^{\alpha\beta}{}_\gamma
              & (\delta_\Wang F_\Wang)^c{}_\alpha \\ \hline\hline
 A4 & c_{12}{}^2 = 2, \, c_{14}{}^4 = 1 &
      \begin{array}{c}
      c^{13}{}_2 = -\frac{b}{a^2}, \, c^{13}{}_3 = \frac{1}{a},\\
      c^{14}{}_4 = \frac{1}{2a}
      \end{array} &
      (\delta_\Wang F_\Wang)^c{}_2 = -\frac{3}{a} \Wang^c{}_2 \\ \hline
 A5 & c_{13}{}^3 = -2, \, c_{14}{}^4 = -1 &
      c^{12}{}_2 = -\frac{1}{a}, \, c^{14}{}_4 = -\frac{1}{2a} &
      \mbox{vanishes} \\ \hline
 B2 & c_{12}{}^2 = 2, \, c_{13}{}^3 = 1 &
      \begin{array}{c}
      c^{13}{}_3 = -\frac{1}{2a}, \, c^{14}{}_2 = \frac{b}{a^2},\\
      c^{14}{}_4 = -\frac{1}{a}
      \end{array} &
      (\delta_\Wang F_\Wang)^c{}_2 = \frac{3}{a} \Wang^c{}_2 \\ \hline
 B3 & \begin{array}{c}
      c_{12}{}^2 = 2, \, c_{13}{}^3 = 1,\\
      c_{14}{}^4 = -1, \, c_{24}{}^3 = 1
      \end{array} &
      \begin{array}{c}
      c^{13}{}_1 = -\frac{1}{a}, \, c^{23}{}_2 = \frac{1}{a},\\
      c^{24}{}_1 = -\frac{1}{a}, \, c^{34}{}_2 = \frac{3b}{a^2},\\
      c^{34}{}_4 = \frac{2}{a}
      \end{array} &
       \mbox{vanishes} \\ \hline
 \end{array}$
 \caption{Calculation of $\delta_W F_W$ for a $G$-invariant connection $W$ in the A4, A5, B2 \& B3 cases}
 \label{YMcalculation}
 \end{center}
 \end{table}

 The invariant metrics displayed in Tables \ref{FelsRenner-classification1} and \ref{FelsRenner-classification2} are used to calculate $c^{\alpha\beta}{}_\gamma$.  Combining this with the Wang maps classified in Table \ref{nonred-homPFB},  we calculate $\delta_W F_W$.  The results from these calculations are displayed in Table \ref{YMcalculation}.  Imposing the Yang--Mills equation $\delta_W F_W = 0$, we obtain:
 
 \begin{theorem} \label{YM-thm} Given the same hypotheses as in Theorem \ref{classification-thm}, suppose that $P_\homo \ra G/K$ admits a $G$-invariant connection.  Then:
 \begin{enumerate}
 \item In the A4 \& B2 cases, there is a unique $G$-invariant Yang--Mills connection, namely the canonical flat connection on the trivial bundle $P=G/K \times H$.
 \item If $G/K$ is of class A5 or B3, {\em all} $G$-invariant connections are Yang--Mills.
 \end{enumerate}
 \end{theorem}

 We now give more concrete geometrical realizations of the $G$-invariant Yang--Mills connections $\omega$ in the A5 \& B3 cases that were derived algebraically.  More precisely, we will: (1) construct global models $G/K$, (2) choose local coordinates on $G/K$, and (3) choose a local section $\tilde\sigma$ of $P_\homo \ra G/K$ and express the local gauge potential $\tilde\sigma^*\omega$ in the local coordinates.

Let $\sigma : U \subset G/K \ra G$ be a local section of $G \ra G/K$.  This induces a local section of $\pi : P_\homo \ra G/K$, namely
 \begin{align*}
   \tilde\sigma : U \ra \pi^{-1}(U), \quad\quad u \mapsto \PFB{\sigma(u)}{e} = L_{\sigma(u)}p.
 \end{align*}
 where $p=\PFB{e}{e}$.  Let $\omega$ be a $G$-invariant connection $P_\homo$ and let $\tilde\omega = \tilde\sigma^*\omega$ be the local gauge potential with respect to $\tilde\sigma$.  Then for $X \in T_u(G/K)$ and using $G$-invariance of $\omega$ and \eqref{omega-W}, we have
 \begin{align}
   \tilde\omega_u(X) &= \omega_{\tilde\sigma(u)}(\tilde\sigma_*(X)) = \omega_p( L_{\sigma(u)^{-1}*} \tilde\sigma_*(X)) = \omega_p( L_{\sigma(u)^{-1}*} (\sigma_*(X),0)^*_p)  \label{omega-explicit}\\
    &= \omega_p( (L_{\sigma(u)^{-1}*}\sigma_*(X),0)^*_p)
    = \Wang( L_{\sigma(u)^{-1}*}\sigma_*(X) ). \nonumber
 \end{align}

 In the analysis of the B3 and A5 cases, we work in the Fels--Renner bases $\{ \FR{i} \}$ (see Section \ref{Fels-Renner-subsection}) and calculate the matrix $L_{\sigma(u)^{-1}*}\sigma_*$ in order to calculate $\tilde\omega_u$. The parameter $\alpha_i$ appearing in the representations below is associated to $v_i$.

 \subsubsection{B3 case} \label{B3-section} Here, $\g = \sa(2,\R) \times \R$.  We have the commutator relations
 \begin{align*}
   \begin{array}{l}
   [\FR{1},\FR{2}] = 2\FR{2},\\{}
   [\FR{1},\FR{3}] = -2\FR{3},\\{}
   [\FR{2},\FR{3}] = \FR{1},
   \end{array} \quad\quad
   \begin{array}{l}
   [\FR{1},\FR{4}] = \FR{4},\\{}
   [\FR{1},\FR{5}] = -\FR{5},\\{}
   [\FR{2},\FR{5}] = \FR{4},\\{}
   [\FR{3},\FR{4}] = \FR{5},
   \end{array}
 \end{align*}
 and $\k = \langle \FR{3}, \FR{5}+\FR{6} \rangle \cong \R^2$.  We have matrix representations
 \begin{align*}
   \g = \l( \begin{array}{cccc}
   \alpha_1 & \alpha_2 & \alpha_4 & 0\\
   \alpha_3 & -\alpha_1 & \alpha_5 & 0\\
   0 & 0 & 0 & 0\\
   0 & 0 & 0 & \alpha_6
   \end{array} \r), \quad
   \k = \l( \begin{array}{cccc}
   0 & 0 & 0 & 0\\
   \varepsilon_1 & 0 & \varepsilon_2 & 0\\
   0 & 0 & 0 & 0\\
   0 & 0 & 0 & \varepsilon_2
   \end{array} \r),
 \end{align*}
 where $\alpha_i, \varepsilon_i \in \R$.  We have connected Lie group models $G$ and $K$ for $\g$ and $\k$:
 \begin{align*}
   G = \l( \begin{array}{cccc}
   a_{11} & a_{12} & b_1 & 0\\
   a_{21} & a_{22} & b_2 & 0\\
   0 & 0 & 1 & 0\\
   0 & 0 & 0 & e^c
   \end{array} \r), \quad
   K = \l( \begin{array}{cccc}
   1 & 0 & 0 & 0\\
   r_1 & 1 & r_2 & 0\\
   0 & 0 & 0 & 0\\
   0 & 0 & 0 & e^{r_2}
   \end{array} \r),
 \end{align*}
 where $a_{11}a_{22}-a_{12}a_{21}=1$ and $b_i,c,r_i \in \R$.  If we take $g \in G$ and $k \in K$,
 \begin{align*}
   gk = \l( \begin{array}{cccc}
   a_{11} + a_{12}r_1 & a_{12} & b_1 + a_{12} r_2 & 0\\
   a_{21} + a_{22}r_1 & a_{22} & b_2 + a_{22} r_2 & 0\\
   0 & 0 & 1 & 0\\
   0 & 0 & 0 & e^{c+r_2}
   \end{array} \r).
 \end{align*}
 The following is a complete set of scalar invariants for the right $K$-action on $G$:
 \begin{align*}
   x_1 = a_{12}, \quad x_2 = a_{22}, \quad x_3 = b_1 - a_{12}c, \quad x_4 = b_2 - a_{22}c.
 \end{align*}
 These may be taken as {\em global} coordinates on the quotient $G/K$.  The only restriction is that $(x_1,x_2)\neq (0,0)$, so $G/K$ is diffeomorphic to $(\R^2 \backslash (0,0)) \times \R^2$.  On the open set $U \subset G/K$ where $x_2 \neq 0$, take the local section of $G \ra G/K$,
 \begin{align*}
   \sigma(x_1,x_2,x_3,x_4) = \l( \begin{array}{cccc}
   \frac{1}{x_2} & x_1 & x_3 & 0\\
   0 & x_2 & x_4 & 0\\
   0 & 0 & 1 & 0\\
   0 & 0 & 0 & 1
   \end{array} \r).
 \end{align*}
 Then for $u = (x_1,x_2,x_3,x_4) \in U$,
 \begin{align*}
   L_{\sigma(u)^{-1}*}\sigma_{*} &=
   \l( \begin{array}{cccc}
   x_2 & -x_1 & x_1 x_4 - x_2 x_3 & 0\\
   0 & \frac{1}{x_2} & -\frac{x_4}{x_2} & 0\\
   0 & 0 & 1 & 0\\
   0 & 0 & 0 & 1
   \end{array} \r)
   \l( \begin{array}{cccc}
   -\frac{dx_2}{x_2{}^2} & dx_1 & dx_3 & 0\\
   0 & dx_2 & dx_4 & 0\\
   0 & 0 & 0 & 0\\
   0 & 0 & 0 & 0
   \end{array} \r)\\
   &= \l( \begin{array}{cccc}
   -\frac{dx_2}{x_2} & x_2 dx_1 - x_1 dx_2 & x_2 dx_3 - x_1 dx_4 & 0\\
   0 & \frac{dx_2}{x_2} & \frac{dx_4}{x_2} & 0\\
   0 & 0 & 0 & 0\\
   0 & 0 & 0 & 0
   \end{array} \r).
 \end{align*}
 Recall that from Table \ref{nonred-homPFB}, $\Wang=0$ on $\langle e_1,e_3,e_4,e_5 \rangle = \langle \FR{1}, \FR{4}, \FR{5}, \FR{3} \rangle$ and $\Wang(\FR{2}) = \Wang(e_2) \in \ker(ad_{\Wang(e_6)}) = \ker(ad_{\homo_*(\FR{5}+\FR{6})})$.  Thus, by \eqref{omega-explicit}, we have the following local gauge potentials
 \begin{align*}
   \tilde\omega_u = (x_2 dx_1 - x_1 dx_2)\otimes y, \qbox{for any} y \in C_\h(\im(\homo_*)),
 \end{align*}
 corresponding to all $G$-invariant Yang--Mills connections on $P_\homo(G/K,H) \ra G/K$.

 \subsubsection{A5 case} \label{A5-section} The unique $G$-invariant (Yang--Mills) connection $\omega$ in this case is the canonical connection.  We outline the construction of a specific model $G/K$ and a local section $\tilde\sigma$ of $P_\homo \ra G/K$ such that $\tilde\sigma^*\omega=0$, i.e. $\omega$ is {\em pure gauge}.

 Here, $\g = \h \rtimes_{\phi} \sl(2,\R)$, where $\h = A^1_{4,9}$.  We have the commutator relations:
 \begin{align*}
   \begin{array}{l}
   [\FR{1},\FR{2}] = 2\FR{2},\\{}
   [\FR{1},\FR{3}] = -2\FR{3},\\{}
   [\FR{2},\FR{3}] = \FR{1},
   \end{array} \quad\quad
   \begin{array}{l}
   [\FR{1},\FR{5}] = -\FR{5},\\{}
   [\FR{1},\FR{6}] = \FR{6},\\{}
   [\FR{2},\FR{5}] = \FR{6},\\{}
   [\FR{3},\FR{6}] = \FR{5},
   \end{array} \quad\quad
   \begin{array}{l}
   [\FR{4},\FR{7}] = 2\FR{4},\\{}
   [\FR{5},\FR{6}] = \FR{4},\\{}
   [\FR{5},\FR{7}] = \FR{5},\\{}
   [\FR{6},\FR{7}] = \FR{6},
   \end{array}
 \end{align*}
 with $\sl(2,\R) = \langle \FR{1},\FR{2},\FR{3} \rangle$,
 $\h = \langle \FR{4},\FR{5},\FR{6},\FR{7} \rangle$, $\phi = ad^\g|_{\h}$.
 The isotropy subalgebra is $\k = \langle w_1,w_2,w_3 \rangle = \langle \FR{1} + \FR{7}, \FR{3} - \FR{4}, \FR{5} \rangle$.
 We have matrix representations:
 \begin{align*}
   &\h = \l( \begin{array}{cccc}
   -2\alpha_7 & -\alpha_5 & \alpha_6 & -2\alpha_4\\
   0 & -\alpha_7 & 0 & \alpha_6 \\
   0 & 0 & -\alpha_7 & \alpha_5\\
   0 & 0 & 0 & 0
   \end{array} \r), \quad \sl(2,\R) = \l( \begin{array}{cc}
   \alpha_1 & \alpha_2\\
   \alpha_3 & -\alpha_1
   \end{array} \r),\\
   &\k = \l( \l( \begin{array}{cccc}
   -2\varepsilon_1 & -\varepsilon_3 & 0 & 2\varepsilon_2\\
   0 & -\varepsilon_1 & 0 & 0 \\
   0 & 0 & -\varepsilon_1 & \varepsilon_3\\
   0 & 0 & 0 & 0
   \end{array} \r),
   \l( \begin{array}{cc}
   \varepsilon_1 & 0\\
   \varepsilon_2 & -\varepsilon_1
   \end{array} \r) \r).
 \end{align*}
 We identify $X \in \sl(2,\R)$ as $diag(0,X,0) \in \sl(4,\R)$ so that $\phi$ is simply the adjoint map in $\gl(4,\R)$.  We recognize $\k$ as the Bianchi V Lie algebra:
 \begin{align*}
   [w_1, w_2] = -2w_2, \quad [w_1, w_3] = -2w_3, \quad [w_2, w_3] = 0.
 \end{align*}

 We use the Lie group $G = H \rtimes_{\tilde\phi} SL(2,\R)$, where
 \begin{align*}
   H &= \l\{ \l( \begin{array}{cccc}
   e^{-2b_7} & -e^{-b_7}b_5 & e^{-b_7}b_6 & -2b_4\\
   0 & e^{-b_7} & 0 & b_6 \\
   0 & 0 & e^{-b_7} & b_5\\
   0 & 0 & 0 & 1
   \end{array} \r) : b_i \in \R \r\},\\
   SL(2,\R) &= \l\{ A = \l( \begin{array}{cc}
   a_{11} & a_{12}\\
   a_{21} & a_{22}
   \end{array} \r) : det(A) = 1 \r\},
 \end{align*}
 and the homomorphism $\tilde\phi : SL(2,\R) \ra Aut(H)$ is simply
 \begin{align*}
   \tilde\phi_A = Ad^H_{diag(1,A,1)}.
 \end{align*}
 The group multiplication law is the standard semi-direct product multiplication
 \begin{align*}
   (h_1,A_1)\cdot_{\tilde\phi} (h_2,A_2) = (h_1\tilde\phi_{A_1}(h_2),A_1 A_2).
 \end{align*}
 The differential of the induced representation $SL(2,\R) \ra Aut(\h)$ is the given representation $\phi : \sl(2,\R) \ra End(\h)$, so that the Lie algebra of $G$ is indeed $\g$.

 A connected, closed Lie subgroup $K$ of $G$ whose Lie algebra is $\k$ is given by
 \begin{align*}
   K = \l\{ \l( \l( \begin{array}{cccc}
   e^{-2r_1} & -r_3 e^{-r_1} & 0 & 2r_2\\
   0 & e^{-r_1} & 0 & 0 \\
   0 & 0 & e^{-r_1} & r_3\\
   0 & 0 & 0 & 0
   \end{array} \r),
   \l( \begin{array}{cc}
   e^{r_1} & 0\\
   r_2 e^{r_1} & e^{-r_1}
   \end{array} \r) \r) : r_i \in \R \r\}.
 \end{align*}

The right $K$-action on $G$ induces the following transformations of the parameters:
 \begin{align}
   &g \cdot_{\tilde\phi} exp(w_1 t): &&
   b_7 \mapsto b_7 + t, \quad a_{11} \mapsto a_{11} e^t, \quad a_{12} \mapsto a_{12} e^{-t}, \label{A5-finite-1}\\
   &&& a_{21} \mapsto a_{21} e^t, \quad a_{22} \mapsto a_{22} e^{-t} \nonumber\\
   &g \cdot_{\tilde\phi} exp(w_2 t):
   && b_4 \mapsto b_4 - te^{-2b_7}, \label{A5-finite-2}\\
   &&& a_{11} \mapsto a_{11} + a_{12} t, \quad a_{21} \mapsto a_{21} + a_{22} t \nonumber\\
   &g \cdot_{\tilde\phi} exp(w_3 t):
   && b_4 \mapsto b_4 + \frac{t}{2} (b_5 a_{12} - b_6 a_{22})e^{-b_7},\label{A5-finite-3}\\
   &&& b_5 \mapsto b_5 + t a_{22}e^{-b_7}, \quad
   b_6 \mapsto b_6 + t a_{12}e^{-b_7} \nonumber
 \end{align}

 \begin{lemma} For the given A5 model $(G,K)$, a complete set of scalar invariants for the right $K$-action on $G$ is:
 \begin{align*}
    (x_1,x_2,x_3,x_4) &= \l( a_{12} e^{b_7}, a_{22} e^{b_7}, b_5 x_1 - b_6 x_2, a_{21} e^{-b_7} + x_2 b_4 - \frac{1}{2} x_3 b_5 \r),
    %\quad x_3 %= (a_{12}b_5 - a_{22}b_6)e^{b_7}
    %    = ,\\
    %    x_4 %&= a_{21} e^{-b_7} + a_{22} b_4  e^{b_7} - \frac{1}{2} b_5( a_{12} b_5 - a_{22} b_6 ) e^{b_7}\\
    %    & = a_{21} e^{-b_7} + x_2 b_4 - \frac{1}{2} x_3 b_5.
 \end{align*}
 where $(x_1,x_2) \neq (0,0)$.  These are global coordinates on the quotient $G/K$.
 \end{lemma}

 These invariants can be derived by solving the system of 3 first order linear PDE arising from the infinitesimal action corresponding to \eqref{A5-finite-1}-\eqref{A5-finite-3}.  This is done using successive applications of the method of characteristics.

 We have that $G/K$ is diffeomorphic to $(\R^2 \backslash (0,0)) \times \R^2$.
 On the open set $U \subset G/K$ where $x_2 \neq 0$, consider the local section of $G \ra G/K$,
 \begin{align*}
   \sigma(x_1,x_2,x_3,x_4) = \l( \l( \begin{array}{cccc}
   1 & 0 & -\frac{x_3}{x_2} & 0\\
   0 & 1 & 0 & -\frac{x_3}{x_2}\\
   0 & 0 & 1 & 0\\
   0 & 0 & 0 & 1
   \end{array} \r),
   \l( \begin{array}{cc}
   \frac{1+x_1 x_4}{x_2} & x_1\\ x_4 & x_2
   \end{array} \r) \r).
 \end{align*}
 Caution must be exercised in the evaluation of $L_{\sigma(u)^{-1}*}\sigma_{*}(\parder{x_i})$.  This is because in general $exp((x,y)t) \neq (exp(xt),exp(yt))$ for $(x,y) \in T_g(G)$ because of the nontrivial action of $SL(2,\R)$ on $H$.  Let $g(t) = (h(t),A(t))$ be a curve in $G$ which passes through the {\em identity} at $t=0$.  Let $b_i(t), a_{ij}(t)$ be the parameter functions appearing in $g(t)$.  Let $g'_i(0) = L_{\sigma(u)^{-1}*}(\sigma_*(\parder{x_i}))$.  Writing $L_{\sigma(u)*}(g'_i(0))$ explicitly in terms of $\dot{b}_i(0), \dot{a}_{ij}(0)$, and equating this to the specific vector $\sigma_*(\parder{x_i}) \in T_{\sigma(u)}(G)$, we can solve for $\dot{b}_i(0), \dot{a}_{ij}(0)$ which yields $g'_i(0)$.  Explicitly, the final results are:
 \begin{align*}
   g'_1(0) &=\l( \mathbb{O}_4,
   \l( \begin{array}{cc}
   x_4 & x_2\\
   -\frac{x_4{}^2}{x_2} & -x_4
   \end{array} \r) \r),\\
 g'_2(0) &=\l( \l( \begin{array}{cccc}
   0 & -\frac{x_3 x_4}{x_2{}^2} & -\frac{x_3}{x_2} & 0\\
   0 & 0 & 0 & -\frac{x_3}{x_2} \\
   0 & 0 & 0 & \frac{x_3 x_4}{x_2{}^2}\\
   0 & 0 & 0 & 0
   \end{array} \r),
   \l( \begin{array}{cc}
   -\frac{1+x_1 x_4}{x_2} & -x_1\\
   \frac{x_4(1+x_1 x_4)}{x_2{}^2} & \frac{1+x_1 x_4}{x_2}
   \end{array} \r) \r),\\
 g'_3(0) &=\l( \l( \begin{array}{cccc}
   0 & -\frac{x_4}{x_2} & -1 & 0\\
   0 & 0 & 0 & -1 \\
   0 & 0 & 0 & \frac{x_4}{x_2}\\
   0 & 0 & 0 & 0
   \end{array} \r),
   \mathbb{O}_2 \r),\\
 g'_4(0) &=\l( \mathbb{O}_4,
   \l( \begin{array}{cc}
    0 & 0\\ \frac{1}{x_2} & 0
   \end{array} \r) \r),
 \end{align*}
 where $\mathbb{O}_4$ and $\mathbb{O}_2$ are zero matrices.  The unique Wang map $\Wang$ in the A5 case (see Table \ref{nonred-homPFB}) is nontrivial only in the $e_1$ and $e_5$ directions, and $\Wang(e_1) = \Wang(e_5)$.  In the Fels--Renner basis, the only nontrivial direction is the $\FR{7}$ direction.  Thus, $\Wang(g'_i(0))=0$ for all $i$, and so $\tilde\omega=0$.

 \section{The principle of symmetric criticality}
 \label{PSC}

 The existence of invariant Yang--Mills connections leads us to consider the validity of a natural principle called the principle of symmetric criticality (PSC) in the context of the bundle of connections over the base manifold.  We review Palais' original formulation of PSC and the local formulation of PSC due to Anderson, Fels \& Torre.  We prove some general facts about PSC and then proceed to apply PSC in the context of the bundle of connections associated to homogeneous PFB over the non-reductive spaces we have considered.  As we shall see, except for the A5 case, these non-reductive spaces provide examples where PSC {\em fails}.

 \subsection{Global formulation of PSC due to Palais}
 Let $G$ act on $M$, and let $\Sigma$ be the set of points fixed by $G$.  The validity of the following ``global'' version of PSC was investigated by Palais \cite{Palais1979}:

 {\it
 For any $G$-invariant function $f : M \ra \R$ and $p \in \Sigma$,
 \begin{align*}
   d( f|_\Sigma )_p = 0 \quad \Rightarrow \quad df_p = 0,
 \end{align*}
 i.e.\ ``critical symmetric points'' are ``symmetric critical points''
 }

 We can think of applying this to a Lagrangian functional, for which the condition of criticality is given by the corresponding Euler--Lagrange equations, but the point of ambiguity here is that the functional is usually defined in terms of an integral over $M$.  This can pose a problem if $M$ is not compact, which is in fact the case for all of the non-reductive examples we have considered.  As such, we describe the following local formulation investigated by Anderson, Fels \& Torre \cite{AndersonFels1997,AFT2001,FelsTorre2002}.

 \subsection{Local formulation of PSC due to Anderson, Fels \& Torre}
 \label{PSC-local}

Given a bundle $E \ra M$, a natural geometric setting for objects used in the calculus of variations is the variational bicomplex $(\Omega^{*,*}(J^\infty(E)),d_H,d_V)$.  We refer the reader to  \cite{Anderson-VB,Anderson1992} for more details on the objects we will sketch below.  In this setting, the exterior derivative $d$ splits into two operators $d=d_H + d_V$, where $d_H$ and $d_V$ indicate differentiation with respect to the base (i.e.\ horizontal) and fiber (i.e.\ vertical) variables respectively.  Field theoretic Lagrangians are naturally represented as horizontal top-degree forms $\homo \in \Omega^{n,0}(J^\infty(E))$.  This interpretation is natural since Lagrangians are objects which one integrates, and their local dependence is on $(x,u,\partial u,...)$ where $x$ are local coordinates on $M$, $u$ are local coordinates along fibers of $E \ra M$, $\partial u$ are local coordinates along the fibers of $J^1 E \ra E$, etc.  The Euler operator applied to a Lagrangian $\lagrangian$ yields the Euler--Lagrange equations $E(\lagrangian) \in \Omega^{n,1}(J^\infty(E))$ determined through the first variational formula:
 \begin{align*}
   d_V \lagrangian = E(\lagrangian) + d_H \eta.
 \end{align*}
A $G$-action on $E \ra M$ by bundle automorphisms naturally lifts to any jet bundle $J^k(E)$ including $J^\infty(E)$ and induces a $G$-action on $\Omega^{*,*}(J^\infty(E))$ via pullback.  In this sense we can speak of $G$-invariant Lagrangians, and $G$-invariant Euler--Lagrange equations - namely, the associated forms are preserved under pullback by any element of $G$.  A well-known theorem states that if $\lagrangian$ is a $G$-invariant Lagrangian, then $E(\lagrangian)$ is $G$-invariant.  (The converse is not true in general.)

 For our purposes, $E \ra M$ will be the (affine) bundle of connections $\mathcal{C}(P) \ra M$ \cite{Garcia1972}.  (Note that up to this point we have been regarding a connection as a section of a subbundle of $T^*P\otimes \h \ra P$.)  We outline the construction of $\mathcal{C}(P) \ra M$.  Given a PFB $P=P(M,H)$, define $\mathcal{C}(P) = J^1 P / H$.  Since $H$ acts freely (and transitively) on $P$, then $H$-invariant sections of $J^1 P \ra P$ are in 1-1 correspondence with sections of $J^1 P / H \ra M$.  Given $\sigma$ an $H$-invariant section of $J^1 P \ra P$, we have $\sigma(p) = j^1_x s^p$ for some local section $s^p$ about $x=\pi(p)$ with $s^p(x)=p$.  Define $H^\sigma_p = (s^p)_* (T_x M)$.  Owing to $H$-invariance of $\sigma$, the distribution $\{ H^\sigma_p \}_{p \in P}$ is $H$-invariant.  It is in fact also complementary to the vertical distribution so that indeed this corresponds to a connection.  Now suppose that $G$ acts on $P \ra M$ by PFB automorphisms.  The left $G$-action and right $H$-action induce actions on $J^1 P$,
 \begin{align*}
 g \cdot (j^1_x s) = j^1_{gx}(g \cdot s), \qquad  (j^1_x s) \cdot h = j^1_x(s \cdot h),
 \end{align*}
 where $(g \cdot s)(x) = gs(g^{-1}x)$ and $(s \cdot h)(x) = s(x)h$.  Since these actions commute, they pass to the quotient $\mathcal{C}(P) = J^1 P / H$.

% In order to study PSC2 for ${\cal C}(P) \ra M$, we will need to understand how the stabilizers $G_x$, $x \in M$, act on the fibers ${\cal C}(P)_x$.  For PFB over homogeneous spaces we can make this action very explicit, owing to complete global classifications of these bundles.
%   Though we have been interpreting connections as sections of (a subbundle of) $T^*P \otimes \h \ra P$, Garcia has   However, owing to $H$-equivariance, a useful result due to Garcia \cite{Garcia1972} allows us to regard connections on $P$ as sections of an affine bundle ${\cal C}(P) \ra M$ called the bundle of connections.  We briefly outline its construction here.

In general, $G$-invariant sections of $E \ra M$ are {\em not} in 1-1 correspondence with sections of $E/G \ra M/G$ since we have isotropy constraints: $\sigma$ is a $G$-invariant section iff $g \sigma(x) = \sigma(gx)$ for all $g \in G$, which implies that
 \begin{align*}
   g\sigma(x) = \sigma(x),\quad \forall g \in G_x \quad \Rightarrow \quad \sigma \in \kappa(E)_x:=(E_x)^{G_x}.
 \end{align*}
 This determines the fibers of a subbundle $\kappa(E)$ of $E$ called the {\em kinematic bundle}.  The $G$-invariant sections of $E \ra M$ necessarily factor through $\kappa(E)$.  Defining $\overline{\kappa(E)} = \kappa(E)/G$ and $\overline{M} = M/G$, there is a 1-1 correspondence between sections of the quotient bundle $\overline{\kappa(E)} \ra \overline{M}$ and $G$-invariant sections of $E \ra M$.
 \begin{align*}
    \xymatrix{\overline{\kappa(E)} \xynor[d] & \kappa(E) \xynor[l]_{q_\kappa} \xynor[r]^i \xynor[d] & E\xynor[d]\\
 \overline{M} & M \xynor[l]^{q_M} \xynor[r]_{id} & M}
 \end{align*}
 Concrete examples of this {\em kinematic reduction}:
 \begin{itemize}
 \item Stationary, spherically symmetric reduction of metrics on $M=\R^2 \times S^2$:  Here the group is $G = \R \times SO(3)$, where the $SO(3)$-factor acts naturally on $S^2$-factor of $M$, and the $\R$-factor represents time and this acts via translations on the first factor of $M$.  This acts lifts naturally to $T^*M \odot T^*M$ and the bundle of metrics over $M$.  In local coordinates, $(t,r,\theta,\phi)$, any metric has 10 components, while any $G$-invariant metric has only 4.  This is a consequence of the isotropy group $SO(2)$ acting on the fibers of the bundle of metrics.
 \item Parametrization of $G$-invariant connections on $P_\homo(G/K,H)$: The kinematic reduction for the bundle of connections $\mathcal{C}(P_\homo) \ra G/K$ is $\overline{\kappa(\mathcal{C}(P_\homo))} \ra pt$.  The single fiber is precisely the space of all Wang maps $\Wang : \g \ra \h$.
 \end{itemize}

Suppose that $G$ acts effectively and semi-regularly on $M$ with $q$-dimensional orbits, so $\bar{M} = M/G$ is $n-q$ dimensional.  Since we are interested in defining a reduced Lagrangian, we wish $G$-invariant top-degree horizontal forms, i.e.\ forms in $\Omega^{n,0}_G(J^\infty(E))$, to be mapped to top-degree horizontal forms on the quotient, i.e.\ forms in $\Omega^{n-q,0}(J^\infty(\overline{\kappa(E)}))$.  This map will be ``natural'' in the sense that we require it to come from a cochain map of variational bicomplexes, i.e.\ it commutes with $d_H$ and $d_V$.
Anderson, Fels \& Torre \cite{AndersonFels1997, AndersonFels2004} investigated the existence of such a cochain map
 \begin{align*}
   \rho_\chi : \Omega^{*,*}_G(J^\infty(E)) \ra \Omega^{*-q,*}(J^\infty(\overline{\kappa(E)})),
 \end{align*}
 which is defined on a $G$-invariant form $\omega \in \Omega^{r,s}_G(J^\infty(E))$ by:
 \begin{enumerate}
 \item Pull back the form by the prolongation of the inclusion map $i : \kappa(E) \ra E$ to a form in $\Omega^{r,s}_G(J^\infty(\kappa(E)))$ (also denoted $\omega$).  This map is essentially a restriction map to the jets of invariant sections.
 \item Form the ($q$-fold) interior product with a nonvanishing $G$-invariant total $q$-chain $\chi$ to produce a $G$-basic form $\chi \lrcorner \omega \in \Omega^{r-q,s}_G(J^\infty(\kappa(E)))$.  If $(x,v,\partial v,...)$ are local coordinates on $J^\infty(\kappa(E))$, then $\chi$ has the form
 \begin{align*}
   \chi = J(x,v,\partial v,...) \, tot(X_1) \wedge ... \wedge tot(X_q)
 \end{align*}
 where $X_i$ are infinitesimal generators corresponding to the $G$-action on $\kappa(E)$, and $tot$ refers to the total prolongation to a vector field on $J^\infty(\kappa(E))$.
 \item Apply a reduction map $\rho$ to $\chi \lrcorner \omega$ to produce a form $\overline{\omega} \in \Omega^{r-q,s}_G(J^\infty(\overline{\kappa(E)}))$.
 \end{enumerate}

   If such chain $\chi$ exists and $\rho_\chi$ is a cochain map, then it is in fact unique (up to scaling) and moreover it is the total prolongation of a $G$-invariant $q$-chain $\chi_M$ on the base manifold $M$
 \begin{align*}
   \chi_M = J(x) X_1 \wedge ... \wedge X_q, \quad X_i \in \g_M
 \end{align*}
 where $\g_M$ are the infinitesimal generators corresponding to the $G$-action on $M$.  Moreover, we will have a cochain map on the $G$-invariant de Rham complex
 \begin{align*}
   \rho_{\chi_M} : \Omega^*_G(M) \ra \Omega^{*-q}(\overline{M})
 \end{align*}
 i.e.\ which commutes with the exterior derivative.  Conversely, the existence of such a chain and cochain map on $M$ yields a chain and cochain map of variational bicomplexes.

 Consequently, to a $G$-invariant Lagrangian $\lagrangian$, there are two associated Euler--Lagrange equations, namely:
 \begin{itemize}
 \item apply the Euler operator directly, OR
 \item apply the Euler operator to the reduced Lagrangian $\tilde\lagrangian = \rho_\chi(\lagrangian)$
 \end{itemize}
 One can then formulate the following local version of PSC:\\

 {\it
 Suppose that $G$ acts on a bundle $E\ra M$ and a cochain map $\rho_\chi$ exists locally.  Then for {\em every} $G$-invariant Lagrangian $\lagrangian$, any local solution $\tilde{s} : \overline{M} \ra \overline{\kappa(E)}$ to $E(\tilde{\lagrangian}) =0$ corresponds to a local ($G$-invariant) solution $s : M \ra E$ to $E(\lagrangian)=0$, i.e.\ $\tilde{s}$ being critical along (symmetric) variations is sufficient for the corresponding section $s$ to be critical along {\em all} variations.
 }\\

 Anderson, Fels \& Torre established the following necessary and sufficient conditions for PSC:
 \begin{enumerate}
 \item PSC1: $H^q( \g, G_x ) \neq 0, \quad \forall x \in M$, where $q=dim(\g) - dim(G_x)$.
 \item PSC2: $({Vert_p(E)}^*)^{G_x} \cap \left[({Vert_p(E)})^{G_x}\right]^0 = 0, \quad \forall p \in \kappa(E)_x$.
 \end{enumerate}
 PSC1 is a condition on the top-degree relative Lie algebra cohomology group \cite{ChevEilen1948}, defined with respect to the exterior derivative \eqref{triv-d} and Lie derivative \eqref{triv-Lie} with $\h = \R$.  It is equivalent to the existence of a nonvanishing $G$-invariant chain $\chi$ for which $\rho_\chi$ is a cochain map of variational bicomplexes.  Moreover, PSC1 only depends on the orbit structure of $G$ on the base manifold $M$ and not the bundle above $M$.  The PSC2 condition is a constraint on how the isotropy groups act on fibers of the given bundle.  (The superscript ``$0$'' refers to the annihilator subspace.)

Before applying PSC to the bundle of connections, we first discuss particular examples where PSC is valid.

 \subsection{PSC is valid if $G$ is compact}

 Anderson \& Fels \cite{AF-private} established the following useful result (proved by taking a basis of $V^*$ (and its dual) adapted to $(V^*)^K$):

 \begin{prop} \label{PSC2-equiv} Let $\rho : K \ra GL(V)$ be a representation. Then PSC2 holds for $\rho$, i.e.\ $(V^*)^K \cap (V^K)^0 = 0$ iff $V^*$ has a $K$-invariant decomposition
 \begin{align*}
   V^* = (V^*)^K \oplus U.
 \end{align*}
 \end{prop}

 We use this result to establish the validity of PSC2 in the compact case.
 
 \begin{cor} \label{PSC2-rho} Suppose that $K$ is compact.  Then PSC2 holds for $\rho : K \ra GL(V)$.
 \end{cor}
 
 \begin{proof} Let $\langle \cdot, \cdot \rangle$ be any (positive-definite) inner product on $V^*$.  Then
 \begin{align*}
   \langle \alpha, \beta \rangle_K = \int_K \langle k\cdot\alpha, k\cdot\beta \rangle \mu_K, \quad\quad \mu_K = \mbox{Haar measure on } K
 \end{align*}
 is a $K$-invariant inner product on $V^*$.  Now $(V^*)^K$ is a $K$-invariant subspace of $V^*$, so its orthogonal complement $U$ with respect to $\langle \cdot, \cdot \rangle_K$ is also $K$-invariant.  Thus, $V^* = (V^*)^K \oplus U$ is a $K$-invariant decomposition, and so PSC2 holds.
 \end{proof}

 \begin{theorem} \label{PSC-cpt} Suppose that $G$ is a compact group acting on $E \ra M$ by bundle automorphisms with connected isotropy groups.  Then PSC is valid provided that $G$-invariant sections of $E \ra M$ exist.
 \end{theorem}

 \begin{proof}
 PSC1: This was given in Remark 4.8 of \cite{AndersonFels1997}. PSC2: Let $x \in M$.  The isotropy subgroup $K=G_x$ is closed in $G$, so is compact.  Let $\rho : K \ra GL(V)$, $k \mapsto k_*$ be the representation of $K$ on $V = Vert_p(E)$ where $p$ is in the fibre above $x$.  By Corollary \ref{PSC2-rho}, PSC2 is valid.
 \end{proof}

 \subsection{Pseudo-Riemannian symmetric spaces}

 Given a symmetric space $G/K$, let $\g = \k \oplus \s$ be the canonical decomposition with
 \begin{align*}
   [\k,\k]\subset \k, \quad [\k,\s]\subset \s, \quad [\s,\s]\subset \k.
 \end{align*}
 A $G$-invariant pseudo-Riemannian metric on $G/K$ corresponds to an $Ad(K)$-inv\-ariant scalar product on $\s$.  Consequently, $Ad : K \ra GL(\s)$ has image in a pseudo-orthogonal group $O(k,\ell)$, where $k+\ell=dim(\s)$.  At the Lie algebra level this implies that for any $x \in \k$, $ad(x) \in \sl(\s)$ is trace-free.

 \begin{theorem} \label{PSC1-sym-space} Any pseudo-Riemannian symmetric space $G/K$ with $K$ connected satisfies PSC1.
 \end{theorem}

 \begin{proof} Since $K$ is connected, it suffices to verify $H^q(\g,\k) = Z^q(\g,\k) / B^q(\g,\k) \neq 0$.  Let $\{ e_\alpha \}_{\alpha=1}^q$ be a basis for $\s$, $\{ e_{\tilde\alpha} \}$ a basis for $\k$, and $\{ \omega^\alpha, \omega^{\tilde\alpha} \}$ the corresponding dual basis.  Since $Z^q(\g,\k)$ is at most one-dimensional, it suffices to show that $Z^q(\g,\k)$ is nontrivial and that $B^q(\g,\k) = 0$.  Now any top-degree relative chain is automatically closed and vanishes on $\k$ so must be a scalar multiple of $\nu = \omega^1 \wedge ... \wedge \omega^q$.
 We have (using \eqref{triv-Lie} with $\h = \R$)
 \begin{align*}
   (\Lieder{e_{\tilde\alpha}} \nu)(e_1,...,e_q) = -\sum_{\beta=1}^q \nu(e_1,...,[e_{\tilde\alpha},e_\beta],...,e_q) = - \sum_{\beta=1}^q c_{\tilde\alpha \beta}{}^\beta = 0,
 \end{align*}
 which vanishes since $ad(\k) \subset \sl(\s)$.  Thus, $\nu$ is $\k$-invariant and $Z^q(\g,\k) \neq 0$.

 Let $\eta \in \Lambda^{q-1}(\g,\k)$.  Then $d\eta \in \Lambda^q(\g,\k)$, so is necessarily a multiple of $\nu$.  We evaluate it on $e_1 \wedge ... \wedge e_q$.  But since $[\s,\s] \subset \k$, then from the formula for $d$ (c.f.\ \eqref{triv-d} with $\h=\R$) and the fact that $\eta$ vanishes on $\k$, we see that $d\eta(e_1,...,e_q)=0$.  Thus, $B^q(\g,\k)=0$.
 \end{proof}

 \begin{remark} The above result also holds for {\em symplectic} symmetric spaces.
 \end{remark}

 In the {\em Riemannian} symmetric space case, the isotropy group $K$ is necessarily {\em compact}, and so PSC2 is automatically satisfied.

 \begin{theorem} \label{PSC-Riem-sym-space}
   Let $G$ be a connected group acting on $M$ with connected isotropy groups and orbits that are Riemannian symmetric spaces.  Then for {\em any} bundle $E \ra M$ with a $G$-action lifting the given one, PSC holds provided that $G$-invariant sections exist.
 \end{theorem}

 We emphasize here that given a purely Lie group-theoretic condition (i.e.\ Riemannian symmetric space) on the orbits on the base manifold $M$, we have PSC holding for {\em any} field theory on $M$ and hence Lagrangian reduction is ``faithful'' for {\em any} choice of Lagrangian for the chosen field theory.  We note further that compactness of $G$ is not required.

 \subsection{Gauge theory on non-reductive pseudo-Riemannian homogeneous spaces}

 Given a PFB $P=P(M,H)$, the Yang--Mills Lagrangian can be defined as a top-degree horizontal form
 in the variational bicomplex associated to the bundle of connections $\mathcal{C}(P) \ra M$.  More precisely, we assume compactness of $H$, a scalar product $\langle \cdot, \cdot \rangle$ on $\Omega^*(M;Ad(P))$ induced from a metric $\mu$ on $M$ and an $Ad(H)$-invariant metric on $H$, and a volume form $\nu$ on $M$, and thus we have
 \begin{align*}
   \lagrangian_{YM} = \langle F_\omega, F_\omega \rangle \nu,
 \end{align*}
 where $F_\omega$ is the curvature associated with the connection $\omega$.

 Let us re-express the PSC conditions in the context of the bundle of connections $\mathcal{C}(P_\homo) \ra G/K$ associated with a homogeneous PFB $P_\homo = P_\homo(G/K,H)$.  What does the set $\mathcal{C}(P_\homo)_{\overline{e}}$ look like?  By $H$-equivariance, it suffices to describe the value of a connection at a single point $p$ in the fibre of $P_\homo \ra G/K$ above $\overline{e}$.  Using the identification $T_p(P_\homo) \cong (\g\times\h)/\hat{\k}$, we have that
 \begin{align*}
 \mathcal{C}(P_\homo)_{\overline{e}} &\cong \{ \omega : (\g\times\h)/\hat{\k} \ra \h \mbox{ linear } \st \omega(\overline{(0,y)}) = -y \}\\
    &\cong \{ \tWang : \g\times\h \ra \h \mbox{ linear } \st \tWang=0 \mbox{ on } \hat{\k} \mbox{ and } \tWang(0,y) = -y \}\\
    &\cong \{ \Wang : \g \ra \h \mbox{ linear } \st \Wang=\homo_* \mbox{ on } \k \}.
 \end{align*}
 For any $c \in \mathcal{C}(P_\homo)_{\overline{e}}$, $Vert_c (\mathcal{C}(P_\homo)) = T_c (\mathcal{C}(P_\homo)_{\overline{e}}) \cong \Lambda^1_\k(\g;\h)$, and this identification is independent of the choice of $c \in \mathcal{C}(P_\homo)_{\overline{e}}$.  Thus, it suffices to check PSC2 for $V=\Lambda^1_\k(\g;\h)$ with respect to the $K$-action \eqref{K-action}.  The subspace $V^K$ corresponds to the space of ``symmetric variations'' about a given invariant connection.

 We have an induced $K$-invariant scalar product $\langle \cdot , \cdot \rangle$ on $V$.  We can reinterpret the PSC2 condition simply in terms of $\langle \cdot , \cdot \rangle$.  Since $\langle \cdot , \cdot \rangle$ is nondegenerate, we have an isomorphism $\phi : V \ra V^*$, namely $x \mapsto \langle x, \cdot\rangle$.  For a subspace $U$ of $V$, define the ``perp'' subspace $U^\perp = \{ x \in V \st \langle x,y \rangle =0, \forall y \in U \}$.  Then $\langle \cdot , \cdot \rangle$ is totally degenerate on $U \cap U^\perp$, i.e.\ $\langle x,y \rangle=0$ for all $x,y \in U \cap U^\perp$.  Now the isomorphism $\phi$ restricts to give an isomorphism $\phi : V^K \ra (V^*)^K$, but moreover restricts to an isomorphism
 \begin{align}
   \phi : V^K \cap (V^K)^\perp \ra (V^*)^K \cap (V^K)^0.
   \label{phi-iso}
 \end{align}
 Why?  If $x \in V^K \cap (V^K)^\perp$, then $\phi(x) \in (V^*)^K$ and for any $y \in V^K$, $\phi(x)(y) = \langle x,y \rangle =0$.  Thus, $\phi(x) \in (V^K)^0$ and the map is well-defined.  It is surjective since if $\eta \in (V^*)^K \cap (V^K)^0$ and $\eta = \phi(x)$ for some $x \in V^K$, then for any $y \in V^K$, $\langle x,y \rangle = \phi(x)(y)=\eta(y)=0$ and so $x \in V^K \cap (V^K)^\perp$.  Thus, comparing the right side of \eqref{phi-iso} with the PSC2 condition and using $\phi$, we have:

 \begin{theorem} \label{PSC-thm} Let $P_\homo = P_\homo(G/K,H)$.  Suppose that $G/K$ admits a $G$-invariant metric, $\h$ admits an $Ad(H)$-invariant inner product, and that $\langle \cdot , \cdot \rangle$ is the induced scalar product on $\Omega^*(G/K; Ad(P_\homo))$.  Then PSC holds for $\mathcal{C}(P_\homo) \ra G/K$ iff
 \begin{enumerate}
 \item[PSC1.] $H^q(\g,K) \neq 0$, where $q = dim(\g) - dim(K)$.
 \item[PSC2.] The induced scalar product $\langle \cdot , \cdot \rangle$ on $V = \Lambda^1_\k(\g;\h)$ is nondegenerate on $V^K$, i.e.\ $V^K \cap (V^K)^\perp = 0$.
\end{enumerate}
 \end{theorem}

 The PSC1 condition of course guarantees that a $G$-invariant volume form on $G/K$ exists.  The $G$-invariant chain $\chi$ is in fact dual to this volume form with respect to the $G$-invariant metric.

 If the bundle $P_\homo \ra M$ is trivial (i.e.\ $\homo$ is trivial) then the $K$-action on $V = \s^* \otimes \h$ is trivial on the $\h$ factor, so nondegeneracy of $\langle \cdot , \cdot \rangle$ on $V^K = (\s^*)^K\otimes \h$ is equivalent to nondegeneracy of the scalar product induced from $\mu$ on $(\s^*)^K$.  But this is equivalent to nondegeneracy of $\mu$ on $\s^K$.  Thus, we can evaluate PSC2 in the majority of the non-reductive cases we have considered.  The only cases that remain are the nontrivial A5 and B3 cases.

 The A5 case is special in the sense that there is a {\em unique} $G$-invariant connection.  Thus, $V^K=0$ and hence PSC2 is trivially satisfied.  Since PSC1 also holds, we have that PSC holds in this case.  The unique $G$-invariant connection in this case is an example of a {\em universal solution}.  Since all symmetric variations are trivial, the unique Wang map is necessarily a critical point of the Euler--Lagrange equations of the {\em any} reduced $G$-invariant Lagrangian.  By PSC, the unique $G$-invariant connection is then a solution of the Euler--Lagrange equations of {\em any} $G$-invariant Lagrangian defined on the bundle of connections $\mathcal{C}(P_\homo) \ra G/K$.

For the B3 case, in the chosen basis (see Table \ref{FelsRenner-classification2}), the space of Wang maps is
 \begin{align}
   \{ \omega^2\otimes f \st f \in \ker(ad_{\homo(e_6)}) \},
 \label{B3-Wang-maps}
 \end{align}
 where $\omega^1,...,\omega^4$ is the dual basis to $e_1,...,e_4 \in \s$.
 Since \eqref{B3-Wang-maps} is a vector space, we can identify it with $V^K$.  The $K$-invariant metric $\mu$ on $\s$ (see Table \ref{FelsRenner-classification2}) has inverse
 \begin{align}
%    \mu = \l(\begin{array}{cccc}
%        0 & 0 & a & 0\\
%        0 & b & 0 & a\\
%        a & 0 & 0 & 0\\
%        0 & a & 0 & 0
%    \end{array} \r), \quad\quad
   B3: \qquad  \tilde\mu = \l(\begin{array}{cccc}
        0 & 0 & \frac{1}{a} & 0\\
        0 & 0 & 0 & \frac{1}{a}\\
        \frac{1}{a} & 0 & 0 & 0\\
        0 & \frac{1}{a} & 0 & -\frac{b}{a^2}
    \end{array} \r) \label{B3-inv-metric}
 \end{align}
 with respect to the basis $\omega^1, ..., \omega^4$.  On $V^K$,
 \begin{align*}
 \langle \omega^2 \otimes f_1, \omega^2 \otimes f_2 \rangle = \tilde\mu(\omega^2,\omega^2)m(f_1,f_2) = 0,
 \end{align*}
 where $f_i \in \ker(ad_{\homo(e_6)})$, i.e.\ $\langle \cdot , \cdot \rangle$ is totally degenerate on $V^K$.  Thus, PSC2 fails.

 \begin{table}[h]
 \begin{align*}
 \begin{array}{|c|cccccccc|} \hline
 G/K & A1 & A2 & A3 & A4 & A5 & B1 & B2 & B3\\ \hline\hline
 PSC1 & \checkmark & \times & \times & \checkmark & \checkmark & \times & \checkmark & \checkmark\\
 PSC2 & \times & \times & \times & \times & \checkmark & \times & \times & \times \\ \hline
 PSC & \times & \times & \times & \times & \checkmark & \times & \times & \times \\ \hline
 \end{array}
 \end{align*}
 \caption{Validity of PSC for $\mathcal{C}(P_\homo) \ra G/K$ for $G/K$ in the Fels--Renner classification}
 \end{table}

 \begin{theorem}
   For the bundle of connections $\mathcal{C}(P_\homo) \ra G/K$ associated to a homogeneous PFB (with compact semi-simple structure group $H$) over a non-reductive pseudo-Riemannian homogeneous space $G/K$ of dimension 4, PSC is valid iff $G/K$ is of type A5.

 In the A5 case, the unique $G$-invariant connection is a universal solution in the sense that it is a solution of the Euler--Lagrange equations associated to any $G$-invariant Lagrangian defined on $\mathcal{C}(P_\homo) \ra G/K$.
 \end{theorem}

 Let us be explicit and illustrate the failure of PSC in the cases A4, B2, and B3.  The validity of PSC1 in these cases implies the existence of a $G$-invariant volume form $\nu$ on $G/K$ and the existence of a nonvanishing $G$-invariant chain $\chi$.  The evaluation $\chi \lrcorner \nu$ yields a $G$-invariant scalar function which is of course a (nonzero) constant on the single orbit.  Without loss of generality, we may assume that this constant is 1.  Thus, the reduced Lagrangian is simply the scalar function
 \begin{align*}
   \overline{\lagrangian}_{YM}[\Wang] = \langle F_\Wang, F_\Wang \rangle
 \end{align*}
 for any Wang map $\Wang$.  Here $\langle \cdot , \cdot \rangle$ refers to the scalar product on $\Lambda^*(\g,K;\h)$.

 In all three cases, the Wang maps satisfy $\Wang=0$ on $\langle e_1,e_3,e_4 \rangle \subset \s$ with $\im(\Wang)$  an abelian subalgebra of $\h$.  The only commutator with component in the $e_2$ direction is $[e_1,e_2] = 2e_2$.  The only nontrivial component of the curvature $F_\Wang$ is:
 \begin{align*}
   F_\Wang(e_1,e_2) = [\Wang(e_1),\Wang(e_2)] - \Wang([e_1,e_2]) = -2\Wang(e_2)
 \end{align*}
 i.e.
 \begin{align*}
   F_\Wang = -2 (\omega^1 \wedge \omega^2)\otimes \Wang(e_2).
 \end{align*}
 The components of the inverses $\tilde\mu$ of the metrics $\mu$ listed in Tables \ref{FelsRenner-classification1} and \ref{FelsRenner-classification2} are
 \begin{align*}
   &A4 : \l( \begin{array}{cccc}
   \frac{1}{2a} & 0 & 0 & 0\\
   0 & 0 & -\frac{1}{a} & 0\\
   0 & -\frac{1}{a} & -\frac{b}{a^2} & 0\\
   0 & 0 & 0 & \frac{1}{a}
   \end{array} \r), \quad
   B2 : \l( \begin{array}{cccc}
   -\frac{1}{2a} & 0 & 0 & 0\\
   0 & 0 & 0 & -\frac{1}{a} \\
   0 & 0 & \frac{1}{a} & 0\\
   0 & -\frac{1}{a} & 0 & -\frac{b}{a^2}
   \end{array} \r),
 \end{align*}
 and for B3 given by \eqref{B3-inv-metric}.  Thus,
 \begin{align*}
   \overline{\lagrangian}_{YM}[\Wang] &= \langle F_\Wang, F_\Wang \rangle  = \frac{1}{2} \cdot 4 \tilde\mu(\omega^1 \wedge \omega^2, \omega^1 \wedge \omega^2)m(\Wang(e_2),\Wang(e_2)) \\
    &= 2\, det\l(\begin{array}{cc} \tilde\mu(\omega^1,\omega^1) & \tilde\mu(\omega^1,\omega^2)\\
    \tilde\mu(\omega^2,\omega^1) & \tilde\mu(\omega^2,\omega^2) \end{array} \r)m(\Wang(e_2),\Wang(e_2)) = 0.
 \end{align*}
 Since $\overline{\lagrangian}_{YM}[\Wang]=0$, then $E(\overline{\lagrangian}_{YM}) =0$.  Consequently, every Wang map satisfies these equations trivially.
 
 For the A4 and B2 cases, this contradicts the fact that there is in fact a {\em unique} $G$-invariant Yang--Mills connection (namely, corresponding to $\Wang(e_2)=0$) and consequently the failure of PSC is clearly illustrated for these cases.  In the B3 case, we have a proper reduction for the Yang--Mills Lagrangian, but this does not contradict the established failure of PSC in this case.  This is because in our formulation of PSC, the validity of PSC requires {\em every} $G$-invariant Lagrangian to reduce properly.  Thus, there should exist a $G$-invariant Lagrangian for which the Euler--Lagrange equations of the reduced Lagrangian do not give correct assertions about the corresponding invariant connections satisfying the Euler--Lagrange equations of the original Lagrangian.  However, at the present time, the author is unaware of an explicit example of such a Lagrangian.

 \section{Conclusions}
 \label{conclusions}

 In this article, we have initiated the study of invariant gauge fields over non-reductive spaces.  From the classification results in Section \ref{non-reductive}, we have seen that the existence of $G$-invariant connections on homogeneous PFB (with compact semi-simple structure group) over these non-reductive spaces places a strong restriction on the topology of these bundles: many of these are necessarily trivial bundles.  For the two nontrivial examples, namely $P_\homo(G/K,H)$ in the A5 or B3 case, we saw that: (1) both are $\homo$-reductive, and (2) all $G$-invariant connections on these bundles are Yang--Mills.  Future topics for investigation include:
 \begin{enumerate}
   \item If a homogeneous PFB $P_\homo(G/K, H) \ra G/K$ admits a $G$-invariant connection, is it necessarily the true that $(\g,\k)$ is $\homo$-reductive?  (We have shown in Lemma \ref{lambda-reductive-lemma} that the converse is always true.) In particular, given the existence of one $G$-invariant connection, is there any notion of a canonical connection on this bundle?
   \item Carry out an analogous program of classifying $G$-invariant Yang--Mills connections over non-reductive spaces in higher dimensions where the topology of the base manifold could be more complicated.  Are all $G$-invariant connections on {\em nontrivial} homogeneous PFB over non-reductive pseudo-Riemannian homogeneous spaces Yang--Mills?
   \item Are there any examples of universal connections on PFB over non-reductive spaces in higher dimensions which are not pure gauge, i.e.\ whose local gauge potentials are never zero?
   \item Laquer \cite{Laquer1984} has investigated stability properties of the Yang--Mills functional about the canonical connection, but in general there is presently no good theoretical understanding of how the second variation of an $G$-invariant Lagrangian reduces under the action of $G$.  More precisely, does positive or negative-definiteness of the Hessian corresponding to a reduced Lagrangian imply positive or negative-definiteness of the Hessian of the original $G$-invariant Lagrangian?
 \end{enumerate}

 \section{Acknowledgements}

 The author is grateful for helpful discussions with Niky Kamran, Mark Fels, and Ian Anderson.  Ian Anderson's {\tt DiffGeom} package ({\tt DifferentialGeometry} in Maple v.11 and later) was very useful for quickly checking some of the Lie algebra computations.

 \bibliography{YMpaper}
 %\bibliographystyle{unsrt}

%    Bibliographies can be prepared with BibTeX using amsplain,
%    amsalpha, or (for "historical" overviews) natbib style.
 \bibliographystyle{amsalpha}
%    Insert the bibliography data here.

\end{document}